\documentclass{sig-alternate}
\pdfoutput=1
\usepackage{latexsym, mathrsfs}
\usepackage{graphicx}
\usepackage[usenames,dvipsnames,svgnames,table]{xcolor}
\usepackage[linesnumbered,ruled,procnumbered]{algorithm2e}
\usepackage{times}
\usepackage{paralist}
\usepackage{multirow}
\usepackage{xspace}
\usepackage[hyphens]{url}
\usepackage{subfig}
\usepackage{hhline}
\usepackage{listings}
\usepackage{setspace}
\usepackage[titletoc,title]{appendix}

\renewcommand*\ttdefault{cmvtt}  
\usepackage[scaled]{beramono}
\usepackage[T1]{fontenc}
\usepackage[utf8]{inputenc}
\newcommand{\mysinglespacing}{
  \setstretch{1}
}
\lstMakeShortInline[columns=fixed]!

\numberwithin{equation}{section}

\newtheorem{lemma}{Lemma}

\newtheorem{example}{Example}
\newtheorem{formulation}{Problem}

\newcommand{\dhub}{{\sc DataHub}\xspace}
\newcommand{\vv}{\mathcal{V}\xspace}
\newcommand{\gcal}{\mathcal{G}\xspace}
\newcommand{\ee}{\mathcal{E}\xspace}

\newcommand{\cc}{\mathcal{C}\xspace}
\newcommand{\rr}{\mathcal{R}\xspace}

\newcommand{\pp}{\mathcal{P}\xspace}

\newcommand{\eat}[1]{}
\newcommand{\papertext}[1]{}
\newcommand{\techreport}[1]{#1}
\newcommand{\stitle}[1]{\vspace{0.5em}\noindent\textbf{#1}}
\renewcommand{\stitle}[1]{\vspace{1em}\noindent\textbf{#1}}

\newcommand{\techreporttext}[1]{}

\newcommand{\agp}[1]{\textcolor{blue}{[Aditya: #1]}}

    {\end{list}}

\makeatletter
\def\@copyrightspace{\relax}
\makeatother

\begin{document}

\title{Principles of Dataset Versioning: \\ Exploring the Recreation/Storage Tradeoff}

\numberofauthors{6}

\author{
\alignauthor
Souvik Bhattacherjee \\
\affaddr{U. of Maryland, College Park}\\
\email{bsouvik@cs.umd.edu}\\
\and
Amit Chavan \\
\affaddr{U. of Maryland, College Park}\\
\email{amitc@cs.umd.edu}\\
\and
Silu Huang \\
\affaddr{U. of Illinois, Urbana-Champaign}\\
\email{shuang86@illinois.edu}\\
\and
Amol Deshpande \\
\affaddr{U. of Maryland, College Park}\\
\email{amol@cs.umd.edu}\\
\and
Aditya Parameswaran \\
\affaddr{U. of Illinois, Urbana-Champaign}\\
\email{adityagp@illinois.edu}\\
}

\maketitle
\begin{abstract}
The relative ease of collaborative data science
and analysis  
has led to a proliferation of many thousands or millions
of {\em versions} of the same datasets in many scientific and commercial
domains, acquired or constructed at various stages of data analysis across
many users, and often over long periods of time.
Managing, storing, and recreating these dataset versions is a non-trivial task.
The fundamental challenge here is the {\em storage-recreation trade-off}: the more storage
we use, the faster it is to recreate or retrieve versions, while the less storage
we use, the slower it is to recreate or retrieve versions.
Despite the fundamental nature of this problem, there has been a surprisingly little amount of 
work on it.
In this paper, we study this trade-off in a principled manner:
we formulate six problems under various settings, 
trading off these quantities in various ways, 
demonstrate that most of the problems are intractable,
and propose a suite of inexpensive heuristics
drawing from techniques in delay-constrained scheduling,
and spanning tree literature, to solve these problems.
We have built a prototype version management system, that aims to serve as 
a foundation to our \dhub system for facilitating collaborative data science~\cite{bhardwaj2014datahub}.
We demonstrate, via extensive experiments, that our proposed heuristics
provide efficient solutions in practical dataset versioning scenarios.
\end{abstract}


\section{Introduction}
The massive quantities of data being generated every day, and the ease
of collaborative data analysis and data science have led to severe issues in management and retrieval of datasets.
We motivate our work with two concrete example scenarios.
\begin{itemize}
\item \ [Intermediate Result Datasets] For most organizations dealing with large volumes of diverse datasets,
a common scenario is that many datasets are repeatedly analyzed in slightly different ways, with the intermediate
results stored for future use. Often, we find that the intermediate results are the same across many
pipelines (e.g., a {\em PageRank} computation on the Web graph is often part of a multi-step workflow).
Often times, the datasets being analyzed might be slightly different (e.g., results of simple transformations or cleaning
operations, or small updates), but are still stored in their entirety. There is currently no way of reducing the amount of stored data in such a scenario:
there is massive redundancy and duplication (this was corroborated by our discussions with a large software company),
and often the computation required to recompute a given version from another one is small enough to not merit storing a new version.
\item \ [Data Science Dataset Versions] In our conversations with a computational biology group, we found that every time
a data scientist wishes to work on a dataset, they make a private copy, perform modifications
via cleansing, normalization, adding new fields or rows, and then store these modified versions
back to a folder shared across the entire group.
Once again there is massive redundancy and duplication across these copies,
and there is a need to minimize these storage costs while keeping these
versions easily retrievable.
\end{itemize}

\vskip 2pt
\noindent
In such scenarios and many others, it is essential to keep track of versions of datasets
and be able to recreate them on demand; 
and at the same time, it is essential to minimize
the storage costs by reducing redundancy and duplication.
The ability to manage a large number of datasets, their versions, and derived datasets, is a
key foundational piece of a system we are building for facilitating collaborative data science,
called \dhub~\cite{bhardwaj2014datahub}.
\dhub
enables users to keep track of datasets and their versions,
represented in the form of a directed {\em version graph} that
encodes derivation relationships, and to
retrieve one or more of the versions for analysis.

In this paper, we focus on the problem of trading off
storage costs and recreation costs in a principled fashion.
Specifically, the problem we address in this paper is:
given a collection of datasets as well as (possibly) a directed version
graph connecting them, minimize the overall storage for storing the datasets
and the recreation costs for retrieving them.
The two goals conflict with each other --- minimizing storage cost typically
leads to increased recreation costs and vice versa.
We illustrate this trade-off via an example.

\begin{figure}[t!]
  \centering
  \includegraphics[width=1\columnwidth]{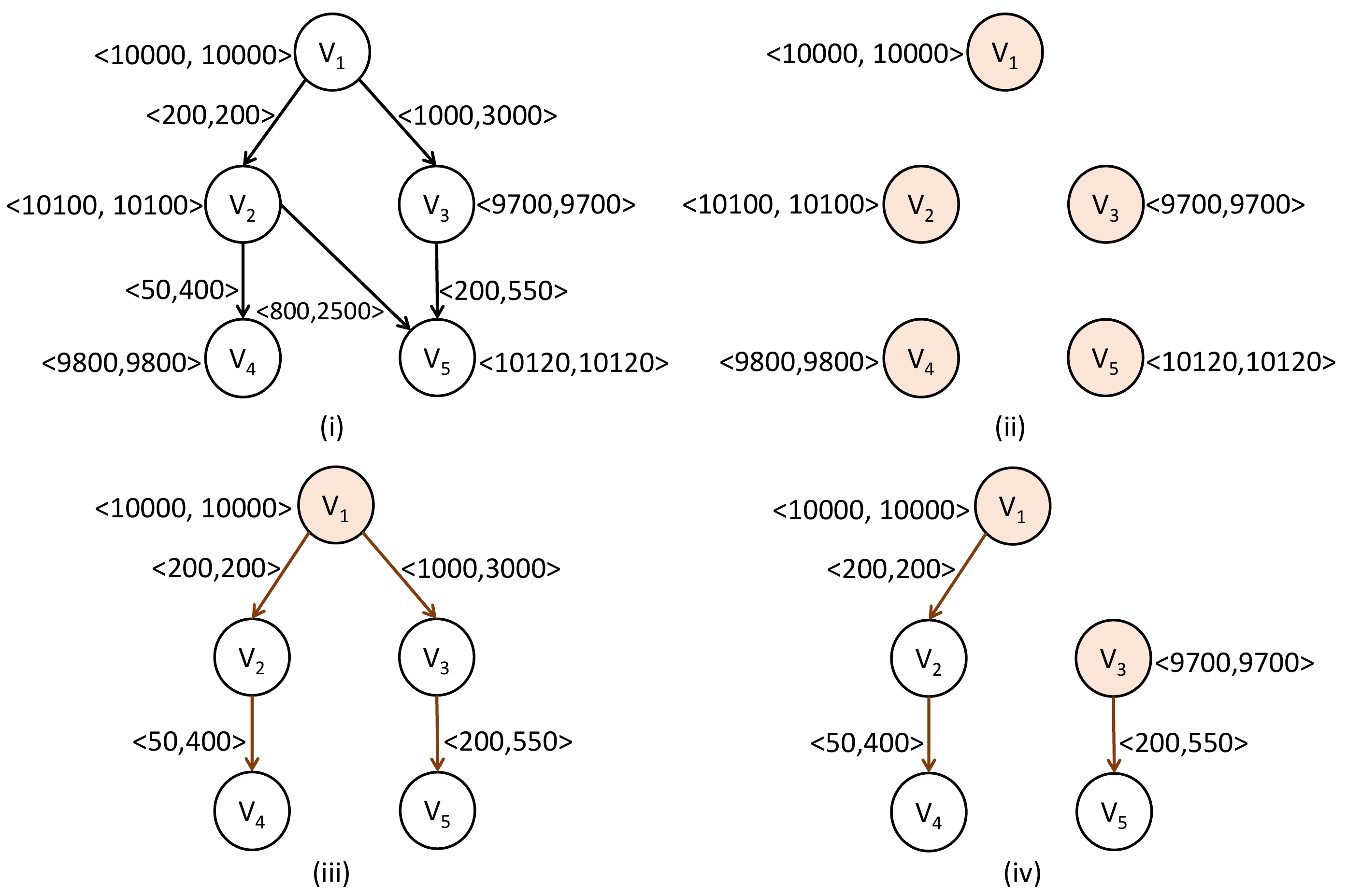}
  \papertext{\vspace{-18pt}}
 \caption{(i) A version graph over 5 datasets -- annotation $\langle a, b\rangle$ indicates
      a storage cost of $a$ and a recreation cost of $b$;
     (ii, iii, iv) three possible storage graphs}
\papertext{     \vspace{-12pt}}
     \label{fig:version_graph}
\end{figure}

\if{0}
\begin{figure}[t!]
  \begin{minipage}[t]{0.5\linewidth}
    \centering
     \includegraphics[width=1\columnwidth]{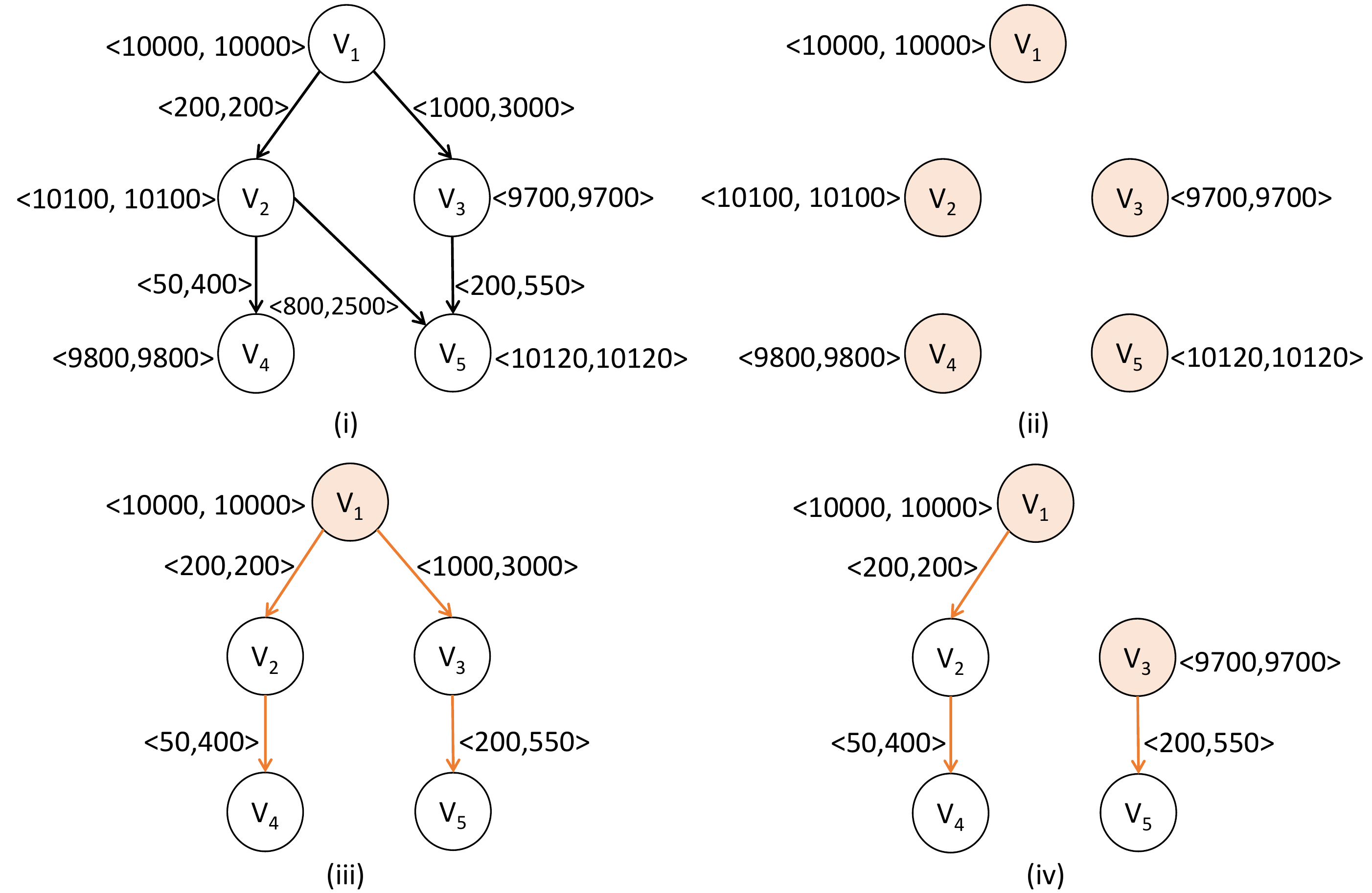}
       \vspace{-12pt}
    (i)
  \end{minipage}
  \begin{minipage}[t]{0.5\linewidth}
    \centering
     \includegraphics[width=1\columnwidth]{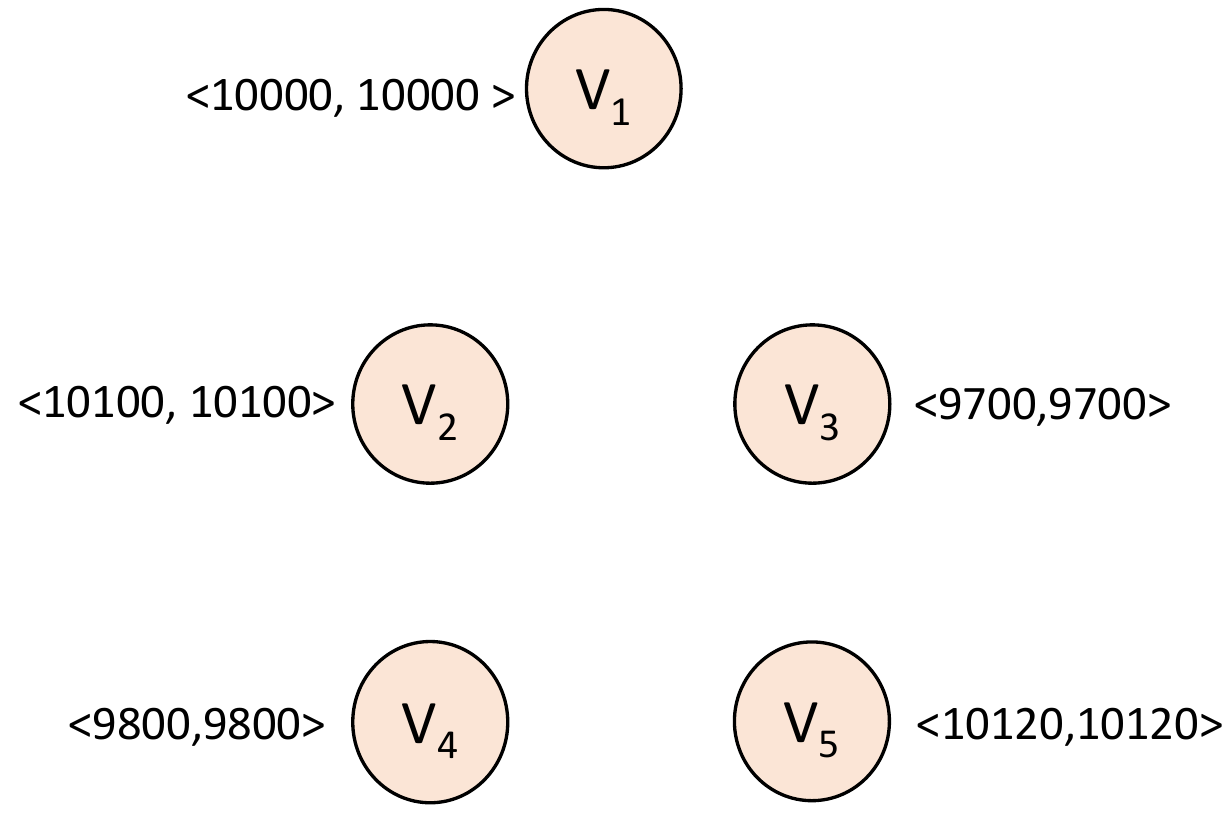}
       \vspace{-12pt}
    (ii)
  \label{fig:extreme1}
  \end{minipage}
  \vspace{5pt}

  \begin{minipage}[t]{0.5\linewidth}
    \centering
     \includegraphics[width=0.9\columnwidth]{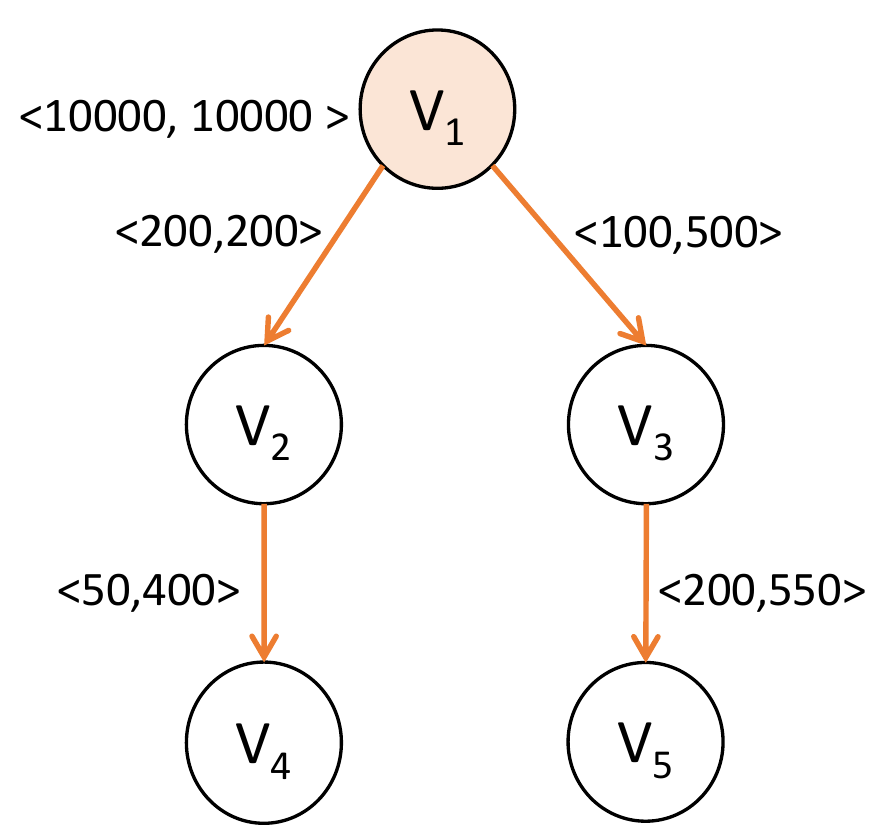}
       \vspace{-15pt}
    (iii)
     \label{fig:extreme2}
  \end{minipage}
  \begin{minipage}[t]{0.5\linewidth}
    \centering
     \includegraphics[width=0.95\columnwidth]{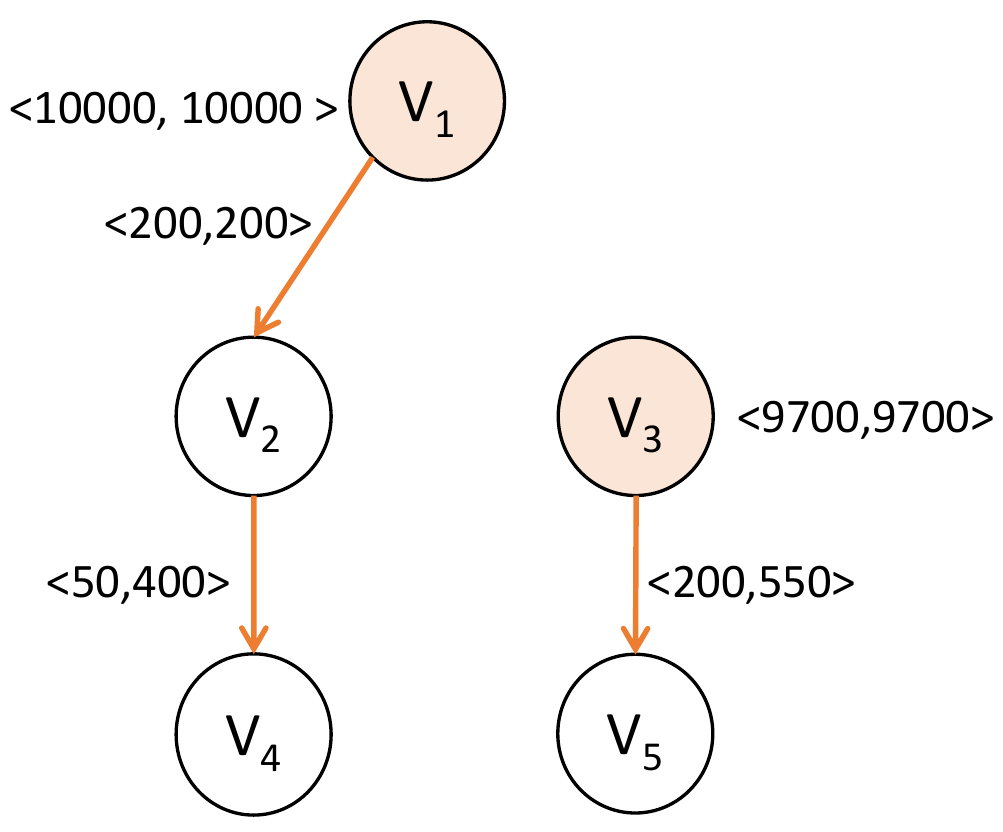}
       \vspace{-15pt}
       (iv)
  \label{fig:storage_graph}
  \end{minipage}
  \caption{(i) A version graph over 5 datasets -- annotation $\langle a, b\rangle$ indicates
      a storage cost of $a$ and a recreation cost of $b$;
     (ii, iii, iv) three possible storage graphs}
     \label{fig:version_graph}
\end{figure}
\fi

\begin{example}
Figure~\ref{fig:version_graph}(i) displays a version graph,
indicating the derivation relationships among 5 versions.
Let $V_1$ be the original dataset.
Say there are two teams collaborating on this dataset:
team 1 modifies $V_1$ to derive $V_2$, while team 2
modifies $V_1$ to derive $V_3$.
Then, $V_2$ and $V_3$ are merged and give $V_5$.
As presented in Figure~\ref{fig:version_graph}, 
$V_1$ is associated with $\langle10000,10000\rangle$, 
indicating that $V_1$'s storage cost and recreation cost are both $10000$ when stored in its entirety
(we note that these two are typically measured in different units -- see the second challenge below); 
the edge $(V_1 \rightarrow V_3)$ is annotated with $\langle 1000,3000\rangle$, 
where $1000$ is the storage cost for $V_3$ when stored as the modification from $V_1$ 
(we call this the {\em delta} of $V_3$ from $V_1$)
and $3000$ is the recreation cost for $V_3$ given $V_1$, i.e,
the time taken to recreate $V_3$ given that
$V_1$ has already been recreated.


One naive solution to store these datasets would be to store all
of them in their entirety (Figure \ref{fig:version_graph} (ii)).
In this case, each version can be retrieved directly
but the total storage cost is rather large, i.e., $10000+10100+9700+9800+10120=49720$.
At the other extreme, only one
version is stored in its entirety while other versions are stored as modifications
or deltas to that version, as shown in Figure~\ref{fig:version_graph} (iii).
The total storage cost here is much smaller ($10000+200+1000+50+200=11450$),
but the recreation cost is large for $V_2,V_3,V_4$ and $V_5$.
For instance, the path $\{(V_1 \rightarrow V_3 \rightarrow V_5)\}$ needs to be accessed
in order to retrieve $V_5$ and the recreation cost is $10000+3000+550=13550>10120$.

Figure \ref{fig:version_graph} (iv) shows an intermediate solution that trades off
increased storage for reduced recreation costs for some version. Here
we store versions $V_1$ and $V_3$ in their entirety and store
modifications to other versions. This solution also exhibits higher
storage cost than solution (ii) but lower than (iii), and still results in
significantly reduced retrieval costs for versions $V_3$ and $V_5$ over (ii).

\end{example}

\noindent
Despite the fundamental nature of the storage-retrieval
problem, there is surprisingly little prior work on
formally analyzing this trade-off and on designing techniques for identifying effective
storage solutions for a given collection of datasets. Version Control Systems (VCS) like Git, SVN,
or Mercurial, despite their popularity, use fairly simple algorithms underneath, and are known to have
significant limitations when managing large datasets~\cite{git-repack,git-comment-fb}.
Much of the prior work in literature focuses on a linear chain of versions, or on minimizing
the storage cost while ignoring the recreation cost (we discuss the related work
in more detail in Section~\ref{sec:related}).

In this paper, we initiate a formal study of the problem of deciding how to jointly store a collection
of dataset versions, provided along with a version or derivation graph. Aside from being able to handle
the scale, both in terms of dataset sizes and the number of versions, there are several other considerations
that make this problem challenging.
\begin{itemize}
  \item Different application scenarios and constraints
  lead to many variations on the basic theme of balancing
  storage and recreation cost (see Table~\ref{table:prob}).
  The variations arise both out of different ways to
  reconcile the conflicting optimization goals, as well as because of the variations
  in how the differences between versions are stored and how versions are reconstructed.
  For example, some mechanisms for constructing differences between versions lead to
  symmetric differences (either version can be recreated from the other version) --- we call
  this the {\em undirected} case. The scenario with asymmetric, one-way differences is referred
 to as {\em directed} case.

  \item Similarly, the relationship between storage and recreation costs leads to significant variations across different settings.
  In some cases the recreation cost is proportional to the storage cost (e.g., if the system bottleneck lies in the I/O cost or network communication),
  but that may not be true when the system bottleneck is CPU computation. This is especially true for sophisticated differencing mechanisms where
  a compact derivation procedure might be known to generate one dataset from another.
  
  \item Another critical issue is that computing deltas for all pairs of versions is typically not feasible. Relying purely on the version
  graph may not be sufficient and significant redundancies across datasets may be missed.
  
  \item Further, in many cases, we may have information about relative {\em access frequencies} indicating the relative likelihood of retrieving
  different datasets. Several baseline algorithms for solving this problem cannot be easily adapted to incorporate such access frequencies.
\end{itemize}

\vskip 3pt
\noindent
We note that the problem described thus far is inherently ``online'' in that 
new datasets and versions are typically being created continuously and are being added to
the system. 
In this paper, we focus on the static, off-line version of this problem and 
focus on formally and completely understanding that version. 
We plan to
address the online version of the problem in the future.
The key contributions of this work are as follows.



\begin{itemize}
  \item We formally define and analyze the dataset versioning problem 
  and consider several variations of the problem that trade off storage cost and
  recreation cost in different manners, under different assumptions about 
  the differencing mechanisms and recreation costs (Section~\ref{sec:probOverview}).
  Table~\ref{table:prob}
  summarizes the problems and our results. We show that most of the variations of this problem are NP-Hard (Section~\ref{sec:complexity}).
  \item We provide two light-weight heuristics: one, when there is a
  constraint on average recreation cost, and one when there is a constraint on
  maximum recreation cost; we also show how we can adapt a prior solution 
  for balancing minimum-spanning trees and shortest path trees for undirected
  graphs (Section~\ref{sec:algorithms}).
  \item We have built a prototype system where we implement the proposed algorithms. 
  We present an extensive experimental evaluation of these algorithms over several
  synthetic and real-world workloads demonstrating the effectiveness of our algorithms 
  at handling large problem sizes (Section \ref{sec:experiments}). 
\end{itemize}

\begin{table*}[t]
\centering
\begin{tabular}{|l||l|l||p{2.3cm}|p{2cm}|p{2cm}|}
  \hline
   & Storage Cost & Recreation Cost& Undirected Case, $\Delta=\Phi$ & Directed Case, $\Delta=\Phi$ & Directed Case, $\Delta \neq \Phi$\\
  \hline
  \hline
  Problem~\ref{prob:minstor} & minimize \{$\mathcal{C}$\} & $\mathcal{R}_i < \infty$, $\forall i$& \multicolumn{3}{c|}{PTime, Minimum Spanning Tree} \\
  \hline
  Problem~\ref{prob:minrec} & $\mathcal{C}< \infty$ & minimize \{$\max\{ \mathcal{R}_i | 1\leq i \leq n$\}\} & \multicolumn{3}{c|}{PTime, Shortest Path Tree} \\ 
  \hline
  Problem~\ref{prob:minsumrec} & $\mathcal{C} \leq \beta$ & minimize \{$\sum_{i=1}^{n} \mathcal{R}_i$\} & NP-hard, & \multicolumn{2}{c|}{NP-hard, LMG Algorithm} \\ 
  \cline{1-3}  \cline{5-6}
  Problem~\ref{prob:minmaxrec} & $\mathcal{C} \leq\beta$ & minimize \{$\max\{ \mathcal{R}_i | 1\leq i \leq n$\}\} & LAST Algorithm$^\dagger$ &\multicolumn{2}{c|}{NP-hard, MP Algorithm} \\
  \hline
  Problem~\ref{prob:minstor-sumrec} & minimize \{$\mathcal{C} $\} & $\sum_{i=1}^{n} \mathcal{R}_i \leq \theta$& NP-hard, &\multicolumn{2}{c|}{NP-hard, LMG Algorithm} \\
   \cline{1-3}  \cline{5-6}
  Problem~\ref{prob:minstor-maxrec} & minimize \{$\mathcal{C}$\} & $\max \{ \mathcal{R}_i| 1\leq i \leq n\} \leq \theta$ & LAST Algorithm$^\dagger$ & \multicolumn{2}{c|}{NP-hard, MP Algorithm} \\
    \hline

\end{tabular}
\caption{Problem Variations With Different Constraints, Objectives and Scenarios.}
\papertext{\vspace{-15pt}}
\label{table:prob}
\end{table*}

\if{0}
\begin{table*}[t]
\centering
\begin{tabular}{|l||l|l||p{2.3cm}|p{2cm}|p{2cm}|}
  \hline
   & Storage Cost & Recreation Cost& Undirected Case, $\Delta=\Phi$ & Directed Case, $\Delta=\Phi$ & Directed Case, $\Delta \neq \Phi$\\
  \hline
  \hline
  Prob~\ref{prob1} & minimize \{$\mathcal{C}$\} & $\mathcal{R}_i < \infty$, $\forall i$& \multicolumn{3}{c|}{PTime, Minimum Spanning Tree} \\
  \hline
  Prob~\ref{prob2} & $\mathcal{C}< \infty$ & minimize \{$\max\{ \mathcal{R}_i | 1\leq i \leq n$\}\} & \multicolumn{3}{c|}{PTime, Shortest Path Tree} \\ 
  \hline
  Prob~\ref{prob3} & $\mathcal{C}< \beta$ & minimize \{$\sum_{i=1}^{n} \mathcal{R}_i$\} & & \multicolumn{2}{c|}{NP-hard, Local move-based Algorithm} \\ 
  \cline{1-3}  \cline{5-6}
  Prob~\ref{prob4} & $\mathcal{C}< \beta$ & minimize \{$\max\{ \mathcal{R}_i | 1\leq i \leq n$\}\} & NP-hard, & NP-hard \agp{? Must be atleast NPH} & NP-hard \\
  \cline{1-3}  \cline{5-6}
  Prob~\ref{prob5} & minimize \{$\mathcal{C} $\} & $\sum_{i=1}^{n} \mathcal{R}_i < \theta$& $(1+2/(\alpha-1), \alpha)$ & \agp{?} & \agp{?} \\
   \cline{1-3}  \cline{5-6}
  Prob~\ref{prob6} & minimize \{$\mathcal{C}$\} & $\max \{ \mathcal{R}_i| 1\leq i \leq n\} < \theta$ & -Approximation$^\dagger$ & \multicolumn{2}{c|}{NP-hard, Prim-based Algorithm} \\
    \hline
  Prob~\ref{prob7} & $\mathcal{C}< \beta$ & minimize \{$\max\{\mathcal{H}_i | 1\leq i \leq n \}$\} &  NP-hard &\multicolumn{2}{c|}{NP-hard, \agp{?}} \\
   \hline
  Prob~\ref{prob8} & minimize \{$\mathcal{C} $\} & $\max\{ \mathcal{H}_i| 1\leq i \leq n\} < \theta$& NP-hard, $\log^{2} n$-approx &\multicolumn{2}{p{5cm}|}{NP-hard, $\log^{3} n$-approximation, Heuristic\agp{list alg no.}} \\
  \hline

\end{tabular}
\caption{Problems With Different Constraints, Objectives and Scenarios. $^\dagger$the total storage cost of resulting graph is within $1+2/(\alpha-1)$-approximation of minimum total storage cost without constraint on recreation cost; while the recreation cost for each version in the resulting graph is within $\alpha$-approximation of minimum recreation cost without constraint on total storage cost.
}
\label{table:prob}
\end{table*}
\fi

\section{Problem Overview}\label{sec:probOverview}

%

In this section, we first introduce essential notations and then present the various problem formulations.
We then present a mapping of the basic problem to a graph-theoretic problem, and also describe
an integer linear program to solve the problem optimally.




\vspace{5pt}
\subsection{Essential Notations and Preliminaries}\label{ssec:notation}
\vspace{-5pt}

\stitle{Version Graph.} We let $\vv = \{V_i\}, i = 1, \ldots, n$ be a collection of versions.
The derivation relationships between versions are represented or captured in the form of a {\em version graph}:
$\gcal(\vv, \ee)$. A directed edge from $V_i$ to $V_j$ in $\gcal(\vv, \ee)$ represents that $V_j$ was derived from $V_i$
(either through an update operation, or through an explicit transformation).
Since branching and merging are permitted in \dhub (admitting collaborative data science),
$\gcal$ is a DAG (directed acyclic graph) instead of a linear chain.
For example, Figure~\ref{fig:version_graph} represents a version graph $\gcal$, where
$V_2$ and $V_3$ are derived from $V_1$ separately, and then merged to form $V_5$.

\stitle{Storage and Recreation.} Given a collection of versions $\vv$, we need to reason about the {\em storage cost},
i.e., the space required to store the versions, and the {\em recreation cost}, i.e., the time taken to recreate
or retrieve the versions. For a version $V_i$, we can either:
\begin{itemize}
\item \underline{Store $V_i$ in its entirety:} in this case, we denote the storage required to record version $V_i$ fully by $\Delta_{i, i}$.
The recreation cost in this case is the time needed to retrieve this recorded version; we denote that by $\Phi_{i, i}$.
A version that is stored in its entirety is said to be {\em materialized}.
\item \underline{Store a ``delta'' from $V_j$:} in this case, we do not store $V_i$ fully; we instead store its modifications from another version $V_j$.
For example, we could record that $V_i$ is just $V_j$ but with the $50$th tuple deleted.
We refer to the information needed to construct version $V_i$ from version $V_j$ as the {\em delta} from $V_j$ to $V_i$.
The algorithm giving us the delta is called a {\em differencing algorithm}.
The storage cost for recording modifications
from $V_j$, i.e., the size the delta, is denoted by $\Delta_{j, i}$. The recreation cost is the
time needed to recreate the recorded version given that $V_j$ has been recreated; this is denoted by $\Phi_{j, i}$.
\end{itemize}

\vskip 2pt
\noindent
Thus the storage and recreation costs can be represented using two matrices $\Delta$ and $\Phi$: the entries along
the diagonal represent the costs for the materialized versions, while
the off-diagonal entries represent the costs for deltas.
From this point forward, we focus our attention on these matrices: they capture all
the relevant information about the versions for managing and retrieving them.

\stitle{Delta Variants.} Notice that by changing the differencing
algorithm, we can produce deltas of various types:
\begin{itemize}
\item for text files, UNIX-style diffs, i.e., line-by-line modifications between versions, are commonly used;
\item we could have a listing of a program, script, SQL query, or command that generates version $V_i$ from $V_j$;
\item for some types of data, an XOR between the two versions can be an appropriate delta; and
\item for tabular data (e.g., relational tables), recording the differences at the cell level is yet another type of delta.
\end{itemize}
Furthermore, the deltas could be stored compressed or uncompressed.
The various delta variants lead to various dimensions of problem that we will describe subsequently.

The reader may be wondering why we need to reason about two matrices $\Delta$ and $\Phi$. In some cases,
the two may be proportional to each other (e.g., if we are using uncompressed UNIX-style diffs).
But in many cases, the storage cost of a delta and the recreation cost of applying that delta can be very
different from each other, especially if the deltas are stored in a compressed fashion.
Furthermore, while the storage cost is more straightforward to account for in that it is proportional to the bytes
required to store the deltas between versions, recreation cost is more complicated: it could
depend on the network bandwidth (if versions or deltas are stored remotely), the I/O bandwidth, and
the computation costs (e.g., if decompression or running of a script is needed).

\begin{example}
Figure~\ref{fig:matrix_phi} shows the matrices $\Delta$ and $\Phi$ based on version graph in Figure~\ref{fig:version_graph}.
The annotation associated with the edge $(V_i, V_j)$ in Figure \ref{fig:version_graph} is essentially
$\langle \Delta_{i,j},\Phi_{i,j} \rangle$, whereas the vertex annotation for $V_i$ is $\langle \Delta_{i,i}, \Phi_{i,i} \rangle$.
If there is no edge from $V_i$ to $V_j$ in the version graph, we have two choices: we can either set the
corresponding $\Delta$ and $\Phi$ entries to ``$-$'' (unknown) (as shown in the figure), or we can explicitly
compute the values of those entries (by running a differencing algorithm).
For instance, $\Delta_{3,2}=1100$ and $\Phi_{3,2}=3200$ are computed explicitly in the figure (the specific
numbers reported here are fictitious and not the result of running any specific algorithm).
\end{example}

\begin{figure}[t!]
  \centering
  \includegraphics[width=1\columnwidth]{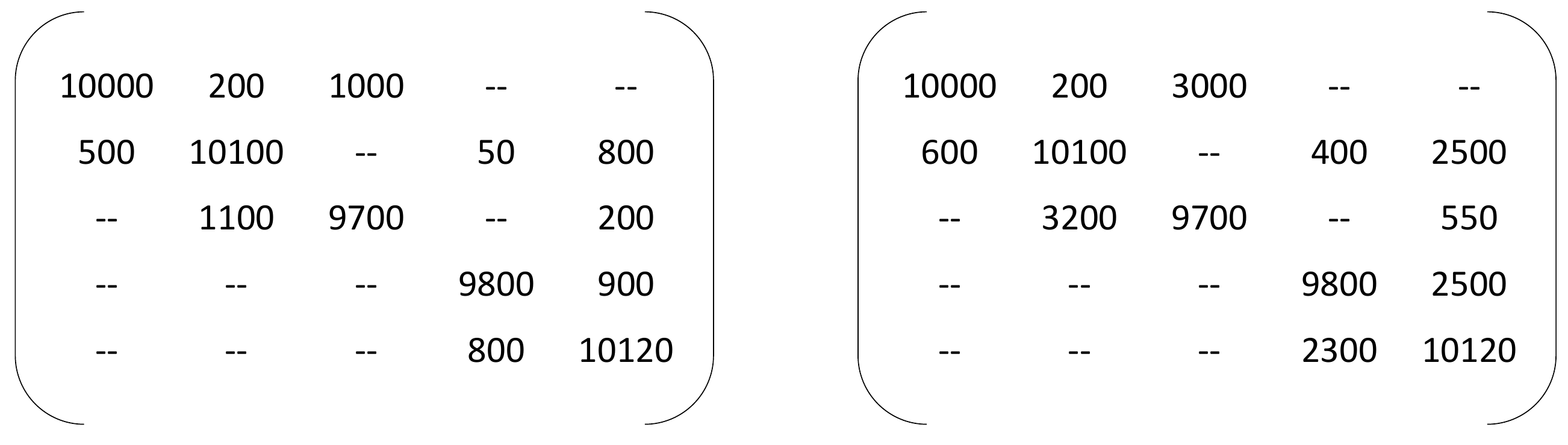}
  \papertext{\vspace{-20pt}}

  (i) $\Delta$ ~~~~~~~~~~~~~~~~~~~~~~~~~~~~~~~~~~~~~~~~~~~~~~~~~~(ii) $\Phi$
  \papertext{\vspace{-2pt}}
 \caption{Matrices corresonding to the example in Figure 1 (with additional entries revealed beyond
 the ones given by version graph)}
 \papertext{\vspace{-12pt}}
     \label{fig:matrix_phi}
\end{figure}

\if{0}
\begin{figure}[t!]
  \begin{minipage}[t]{0.5\linewidth}
    \centering
     \includegraphics[width=0.95\columnwidth,height=3cm]{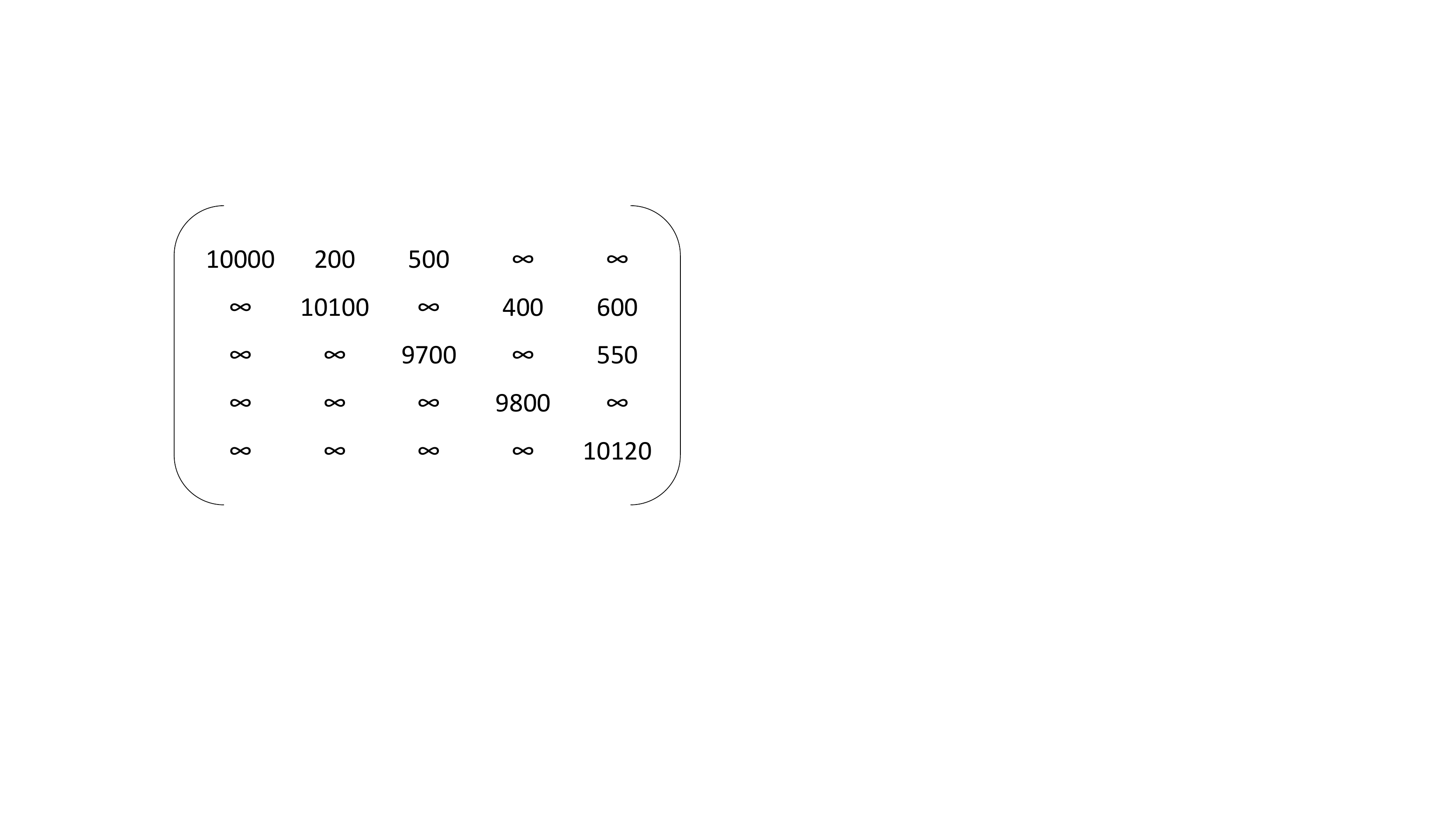}
  \caption{Matrix $\Delta$ }
  \label{fig:matrix_phi}
  \end{minipage}
  \begin{minipage}[t]{0.5\linewidth}
    \centering
     \includegraphics[width=0.95\columnwidth,height=3cm]{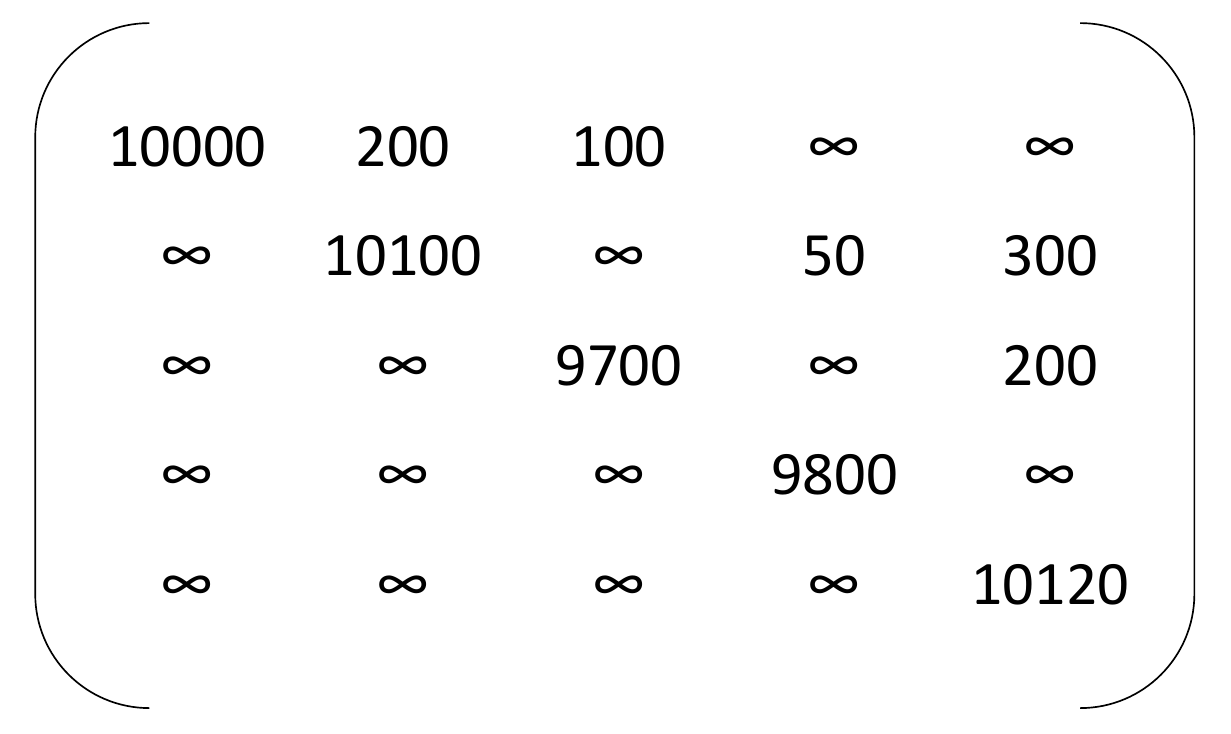}
  \caption{Matrix $\Phi$ } 
  \label{fig:matrix_delta}
  \end{minipage}
\end{figure}
\fi

\papertext{\vspace{-3pt}}
\stitle{Discussion.}
Before moving on to formally defining the basic optimization problem, we note several
complications that present unique challenges in this scenario.
\begin{itemize}
\item {\em Revealing entries in the matrix:}
Ideally, we would like to compute all pairwise
$\Delta$ and $\Phi$ entries, so that we do not miss any significant redundancies among versions
that are far from each other in the version graph.
However when the number of versions, denoted $n$, is large,
computing all those entries can be very expensive (and typically infeasible), since
this means computing deltas between all pairs of versions.
Thus, we must reason with incomplete $\Delta$ and $\Phi$ matrices.
Given a version graph $\gcal$, one option is to restrict our deltas to correspond to actual edges in the
version graph; another option is to restrict our deltas to be between ``close by'' versions, with the understanding
that versions close to each other in the version graph are more likely to be similar. Prior work has also
suggested mechanisms (e.g., based on hashing) to find versions that are close to each other~\cite{Douglis2003}.
We assume that some mechanism to choose which deltas to reveal is provided to us.

\item {\em Multiple ``delta'' mechanisms:} Given a pair of versions $(V_i, V_j)$, there could be
many ways of maintaining a  delta between them, with different $\Delta_{i, j}, \Phi_{i, j}$ costs.
For example, we can store a program used to derive $V_j$ from $V_i$, which could take longer to run
(i.e., the recreation cost is higher) but is more compact (i.e., storage cost is lower),
or explicitly store the UNIX-style diffs between the two versions,
with lower recreation costs but higher storage costs.
For simplicity, we pick one delta mechanism: thus the matrices $\Delta, \Phi$ just have one entry
per $(i, j)$ pair. Our techniques also apply to the more general scenario with small modifications.

\item {\em Branches:} Both branching and merging are common in collaborative analysis, making the version graph a directed acyclic graph.
In this paper, we assume each version is either stored in its entirety or stored as a delta from a single other version, even if it is derived
from two different datasets. Although it may be more efficient to allow a version to be stored as a delta from two other versions in some cases,
representing such a storage solution requires more complex constructs and both the problems of finding an optimal storage solution
for a given problem instance and retrieving a specific version become much more complicated. We plan to further study such solutions in future.

\if{0}
\item {\em Steiner Node:} Consider the following scenario. Suppose we have 3 versions, $V_1, V_2, V_3$ share common portion $V'$
together with some extra portions $\delta_1, \delta_2$ and $\delta_3$ respectively, i.e.,
$V_1=V'+\delta_1, V_2=V'+\delta_2, V_3=V'+\delta_3$
and $\Phi_{1,2}=\delta_1+\delta_2, \Phi_{1,3}=\delta_1+\delta_3, \Phi_{2,3}=\delta_2+\delta_3$.
If creating a Steiner node $V'$ and storing $V_1, V_2, V_3$ as the modification from $V'$, the total storage cost will be $V'+\delta_1+\delta_2+\delta_3$.
Instead, if we store $V_1$ in its entirety(similar analysis can be done when storing $V_2$ in its entirety) and store $V_2, V_3$ as the modification from $V_1$, the total storage cost will be $(V'+\delta_1)+(\delta_1+\delta_2)+(\delta_1+\delta_3)$, which is more expensive than that with a Steiner node.
Though adding steiner nodes can help reduce storage cost in some cases, when and where to add the Steiner nodes pose a great challenge.
\fi
\end{itemize}

\stitle{Matrix Properties and Problem Dimensions.} The storage cost matrix $\Delta$ may be symmetric or asymmetric
depending on the specific differencing mechanism used for constructing deltas. For example, the XOR differencing
function results in a symmetric $\Delta$ matrix since the delta from a version $V_i$ to $V_j$ is identical to
the delta from $V_j$ to $V_i$. UNIX-style diffs where line-by-line modifications are listed can either be two-way
(symmetric) or one-way (asymmetric). The asymmetry may be quite large. For instance, it may be possible
to represent the delta from $V_i$ to $V_j$ using a command like: \emph{delete all tuples with age > 60}, very
compactly. However, the reverse delta from $V_j$ to $V_i$ is likely to be quite large, since all the tuples
that were deleted from $V_i$ would be a part of that delta. In this paper, we consider both these scenarios.
We refer to the scenario where $\Delta$ is symmetric
and $\Delta$ is asymmetric as the undirected case and directed case, respectively.

A second issue is the relationship between $\Phi$ and $\Delta$. In many scenarios, it may be reasonable
to assume that $\Phi$ is proportional to $\Delta$. This is generally true for deltas that contain detailed
line-by-line or cell-by-cell differences. It is also true if the system bottleneck is network communication
or I/O cost. In a large number of cases, however, it may be more appropriate to treat them as independent
quantities with no overt or known relationship.
For the proportional case, we assume that the proportionality constant is 1 (i.e., $\Phi = \Delta$); the
problem statements, algorithms and guarantees are unaffected by having a constant proportionality factor.
The other case is denoted by $\Phi \neq \Delta$.


This leads us to identify three distinct cases with significantly diverse properties:
(1) \textbf{Scenario 1}: Undirected case, $\Phi=\Delta$;
(2) \textbf{Scenario 2}: Directed case, $\Phi=\Delta$; and
(3) \textbf{Scenario 3}: Directed case, $\Phi \neq \Delta$.
%
%

\stitle{Objective and Optimization Metrics.} Given $\Delta, \Phi$, our goal is to find a good storage
solution, i.e., we need to decide which versions to materialize and which versions to store
as deltas from other versions. Let $\pp = \{(i_1, j_1), (i_2, j_2), ...\}$ denote a storage solution.
$i_k = j_k$ indicates that the version $V_{i_k}$ is materialized (i.e., stored explicitly in its entirety),
whereas a pair $(i_k, j_k), i_k \ne j_k$ indicates that we store a delta from $V_{i_k}$ to $V_{j_k}$.

We require any solution we consider to be a {\em valid} solution, where it is possible to reconstruct
any of the original versions. More formally, $\pp$ is considered a {\em valid} solution if and only if
for every version $V_i$, there exists a sequence of distinct versions $V_{l_1}, ..., V_{l_k} = V_i$ such
that $(i_{l_1}, i_{l_1}), (i_{l_1}, i_{l_2}), (i_{l_2}, i_{l_3}), ..., (i_{l_{k-1}}, i_{l_k})$ are
contained in $\pp$ (in other words, there is a version $V_{l_1}$ that can be materialized and can be
used to recreate $V_i$ through a chain of deltas).

We can now formally define the optimization goals:
\begin{itemize}
\item {\em Total Storage Cost} (denoted ${\cal C}$): The total storage cost for a solution $\pp$ is simply the storage cost necessary
to store all the materialized versions and the deltas: $\cc = \sum_{(i, j) \in \pp}{\Delta_{i, j}}$.
\item {\em Recreation Cost for $V_i$} (denoted $\rr_i$): Let $V_{l_1}, ..., V_{l_k} = V_i$ denote a sequence
that can be used to reconstruct $V_i$. The cost of recreating $V_i$ using that sequence is:
$\Phi_{l_1, l_1} + \Phi_{l_1, l_2} + ... + \Phi_{l_{k-1}, l_k}$. The recreation cost for $V_i$ is the minimum
of these quantities over all sequences that can be used to recreate $V_i$.


\end{itemize}

\stitle{Problem Formulations.} We now state the problem formulations that we consider in this paper, starting
with two base cases that represent two extreme points in the spectrum of possible problems.

\papertext{\vspace{-5pt}}
\begin{formulation}[Minimizing Storage] \label{prob1} \label{prob:minstor}
Given $\Delta, \Phi$, find a\\ valid solution $\pp$ such that $\mathcal{C}$ is minimized.
\end{formulation}

\papertext{\vspace{-10pt}}
\begin{formulation}[Minimizing Recreation]\label{prob2} \label{prob:minrec}
Given $\Delta, \Phi$, identify a valid solution $\pp$ such that $\forall i, R_i$ is minimized.
\end{formulation}

The above two formulations minimize either the storage cost or the recreation cost, without worrying about the other. It may
appear that the second formulation is not well-defined and we should instead aim to minimize the average recreation cost across
all versions. However, the (simple) solution that minimizes average recreation cost also naturally minimizes $\mathcal{R}_i$ for each version.

In the next two formulations, we want to minimize (a) the sum of recreation costs over all versions ($\sum_i \mathcal{R}_i$),
(b) the max recreation cost across all versions ($\max_i \mathcal{R}_i$),
under the constraint that total storage cost $\mathcal{C}$ is smaller than some threshold $\beta$.
These problems are relevant when the storage budget is limited. 

\papertext{\vspace{-5pt}}
\begin{formulation}[MinSum Recreation]\label{prob3} \label{prob:minsumrec}
Given $\Delta, \Phi$ and a th- reshold $\beta$, identify $\pp$ such that $\mathcal{C}\leq \beta$, and $\sum_{i} \mathcal{R}_i$ is minimized.
\end{formulation}

\papertext{\vspace{-12pt}}
\begin{formulation}[MinMax Recreation]\label{prob4} \label{prob:minmaxrec}
Given $\Delta, \Phi$ and a th- reshold $\beta$, identify $\pp$ such that $\mathcal{C}\leq \beta$, and $\max_{i} \mathcal{R}_i$ is minimized.
\end{formulation}


The next two formulations seek to instead minimize the total
storage cost $\mathcal{C}$ given a constraint on the sum of recreation costs or max recreation cost.
These problems are relevant when we want to reduce the storage cost, but must satisfy some constraints on the recreation costs.

\papertext{\vspace{-5pt}}
\begin{formulation}[Minimizing Storage(Sum Recreation)]\label{prob5}\label{prob:minstor-sumrec}
Given $\Delta, \Phi$ and a threshold $\theta$, identify $\pp$ such that $\sum_{i} \mathcal{R}_i \leq \theta$, and $\mathcal{C}$ is minimized.
\end{formulation}

\papertext{\vspace{-12pt}}
\begin{formulation}[Minimizing Storage(Max Recreation)]\label{prob6}\label{prob:minstor-maxrec}
Given $\Delta, \Phi$ and a threshold $\theta$, identify $\pp$ such that $\max_i \mathcal{R}_i \leq \theta$, and $\mathcal{C}$ is minimized.
\end{formulation}


\subsection{Mapping to Graph Formulation}\label{ssec:graph}
In this section, we'll map our problem into a graph problem,
that will help us to adopt and modify algorithms from well-studied problems
such as minimum spanning tree construction
and delay-constrained scheduling.
Given the matrices $\Delta$ and $\Phi$,
we can construct a directed, edge-weighted graph $G=(V,E)$ representing the relationship among different versions as follows.
For each version $V_i$, we create a vertex $V_i$ in $G$.
In addition, we create a dummy vertex $V_0$ in $G$.
For each $V_i$, we add an edge $V_0 \rightarrow V_i$, and assign its
edge-weight as a tuple $\langle \Delta_{i,i}, \Phi_{i,i} \rangle$.
Next, for each $\Delta_{i,j}\neq \infty$, we
add an edge $V_i \rightarrow V_j$ with edge-weight
$\langle \Delta_{i,j}, \Phi_{i,j} \rangle$.

The resulting graph $G$ is similar to the original version graph, but with
several important differences. An edge in the version graph indicates a derivation
relationship, whereas an edge in $G$ simply indicates that it is possible to recreate
the target version using the source version and the associated edge delta (in fact,
ideally $G$ is a complete graph).
Unlike the version graph, $G$ may contain cycles, and it also contains the
special dummy vertex $V_0$.
Additionally, in the version graph, if a version $V_i$ has multiple in-edges,
it is the result of a user/application merging changes from multiple versions into $V_i$.
However, multiple in-edges in $G$ capture the multiple choices that we have in recreating $V_i$ from some other versions.

Given graph $G=(V,E)$, the goal of each of our problems
is to identify a storage graph $G_s=(V_s,E_s)$,
a subset of $G$, favorably balancing total storage cost
and the recreation cost for each version.
Implicitly, we will store all versions and deltas
corresponding to edges in this storage graph.
(We explain this in the context of the example below.)
We say a storage graph $G_s$ is
\emph{feasible} for a given problem if (a) each version can be
recreated based on the information contained or stored in $G_s$,
(b) the recreation cost or the total storage
cost meets the constraint listed in each problem.

\begin{figure}[t!]
  \begin{minipage}[t]{0.5\linewidth}
    \centering
     \includegraphics[width=0.9\columnwidth,height=3.5cm]{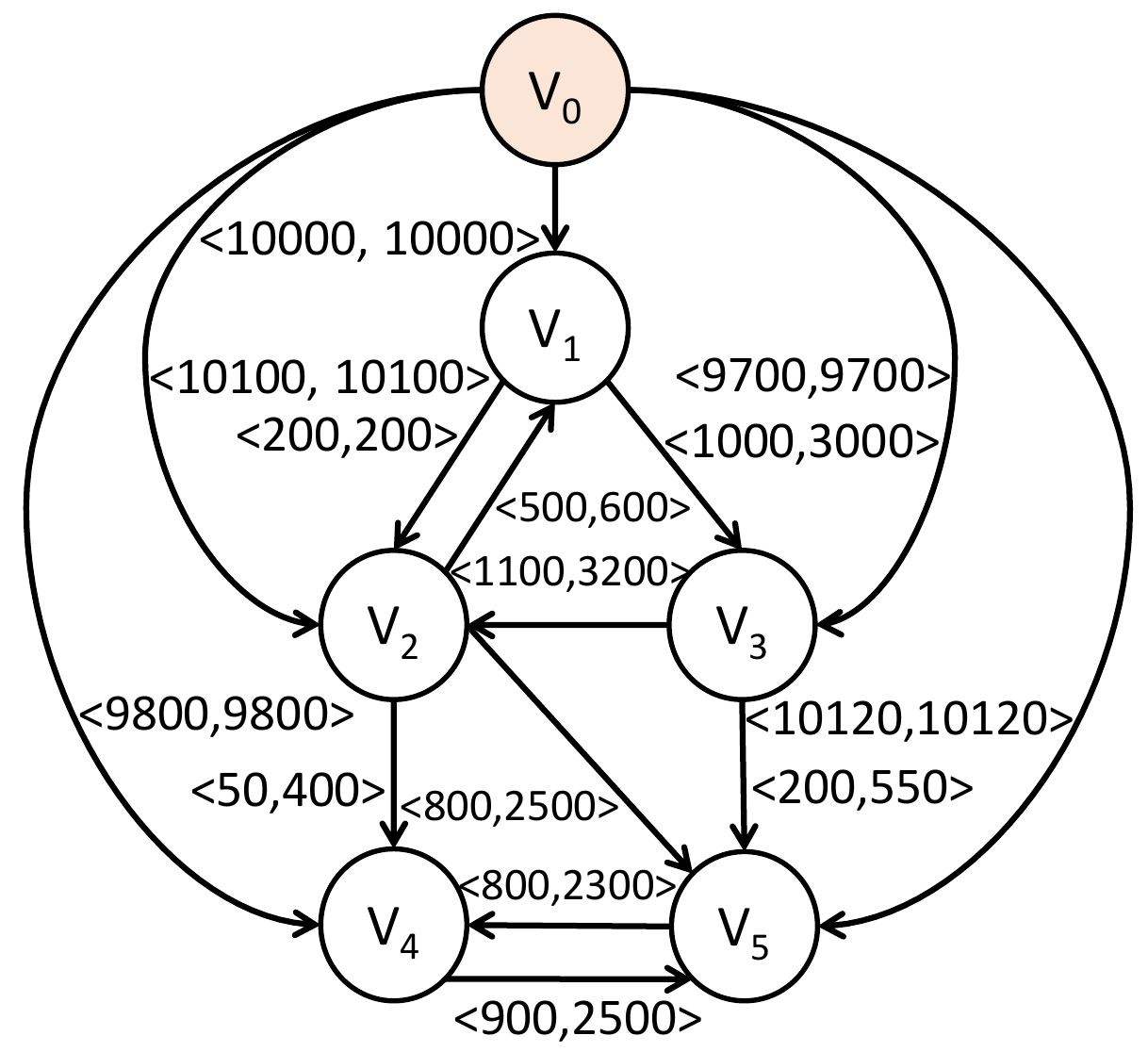}
     \papertext{\vspace{-8pt}}
    \caption{Graph $G$ } 
     \label{fig:dummy_graph}
  \end{minipage}
  \begin{minipage}[t]{0.5\linewidth}
    \centering
     \includegraphics[width=0.65\columnwidth,height=3.5cm]{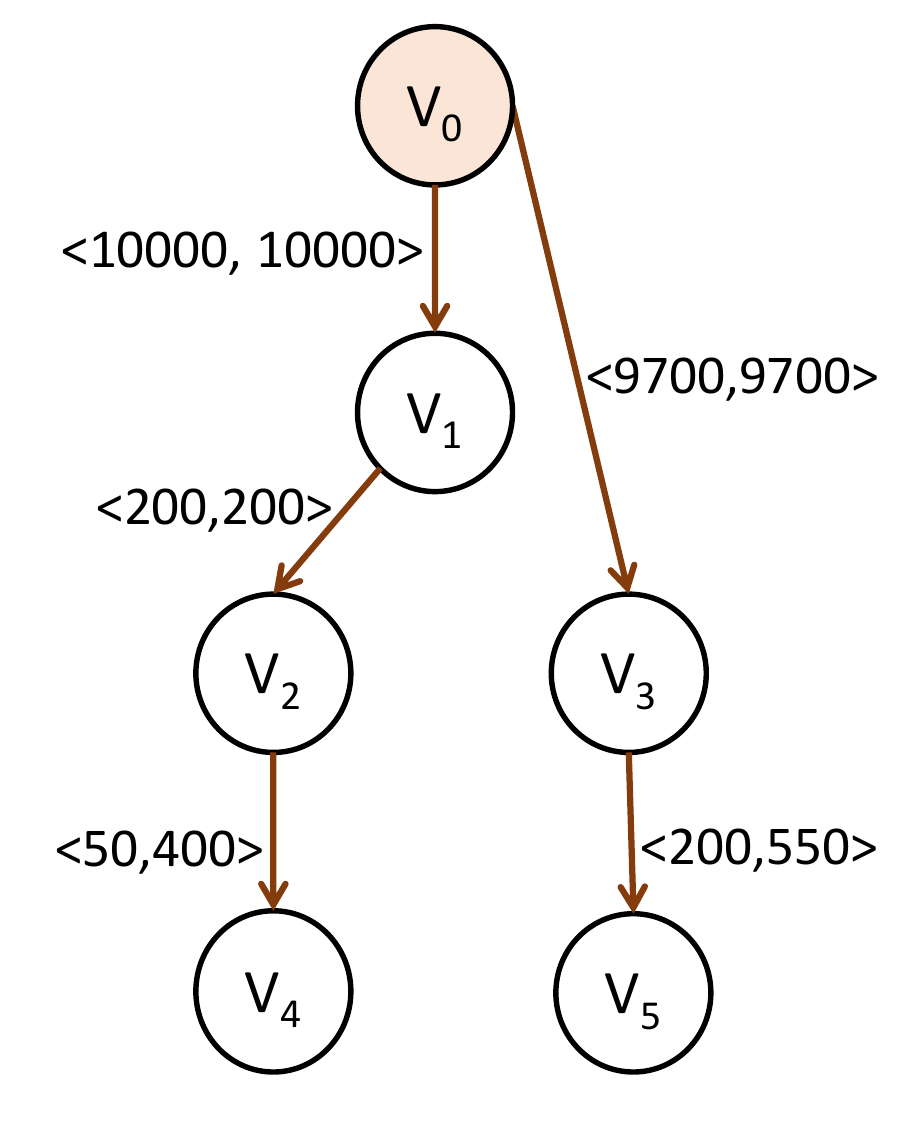}
     \papertext{\vspace{-8pt}}
  \caption{Storage Graph $G_s$} 
  \label{fig:storage_G}
  \end{minipage}
  \papertext{\vspace{-15pt}}
\end{figure}

\papertext{\vspace{-5pt}}
\begin{example}
Given matrix $\Delta$ and $\Phi$ in
Figure~\ref{fig:matrix_phi}(i) and ~\ref{fig:matrix_phi}(ii),
the corresponding graph $G$ is shown in Figure \ref{fig:dummy_graph}.
Every version is reachable from $V_0$.
For example, edge $(V_0,V_1)$ is weighted with $\langle \Delta_{1,1}, \Phi_{1,1}\rangle=\langle 10000, 10000 \rangle$;
edge $\langle V_3, V_5 \rangle$ is weighted with $\langle \Delta_{3,5}, \Phi_{3,5} \rangle=\langle 800, 2500\rangle$.
Figure~\ref{fig:storage_G} is a
feasible storage graph given $G$ in Figure~\ref{fig:dummy_graph},
where $V_1$ and $V_3$ are materialized (since
the edges from $V_0$ to $V_1$ and $V_3$ are present)
while $V_2, V_4$ and $V_5$ are stored as modifications from other versions.
\end{example}
\papertext{\vspace{-5pt}}

After mapping our problem into a graph setting, we have the following lemma.

\papertext{\vspace{-10pt}}
\begin{lemma} \label{MST_SP}
The optimal storage graph $G_s=(V_s,E_s)$ for all 6 problems listed above must be a spanning tree $T$ rooted at dummy vertex $V_0$ in graph $G$.
\papertext{\footnote{We refer the reader to the extended version for proofs.}}
\end{lemma}

\techreport{
\begin{proof}
Recall that a spanning tree of a graph $G(V,E)$ is a subgraph of $G$ that (i) includes all vertices of $G$, (ii) is connected, i.e., every vertex is reachable from every other vertex, and (iii) has no cycles.
Any $G_s$ must satisfy (i) and (ii) in order to ensure that a version $V_i$ can be recreated from $V_0$ by following the path from $V_0$ to $V_i$.
Conversely, if a subgraph satisfies (i) and (ii), it is a valid $G_s$ according to our definition above.
Regarding (iii), presence of a cycle creates redundancy in $G_s$.
Formally, given any subgraph that satisfies (i) and (ii), we can arbitrarily delete one from each of its cycle until the subgraph is cycle free, while preserving (i) and (ii).
\end{proof}
}

\if{0}
\begin{proof}
We first prove that there exists one and only one in-coming edge for each vertex $V_i$ in the optimal storage graph $G_s$, where $i\neq 0$.

It is intuitive that there must exist at least one in-coming edge for each vertex $V_i, i\neq 0$. Otherwise, this vertex is not reachable from $V_0$, making the recreation cost or recreation hop $\infty$.

Next, let us prove that for each vertex $V_i$, there is at most one in-coming edge in optimal storage graph $G_s$.
Suppose there are two or more in-coming edges for $V_i$ in $G_s$, we first find a shortest path weighted with recreation cost(or hop) from the dummy vertex $V_0$ to $V_i$.
Then we can simply delete all other in-coming edges of $V_i$ except the one in the shortest path from $V_0$.
In this resulting spanning tree, the recreation cost(or hop) for each vertex $V_i$ maintains the same,
while the total storage cost is reduced. Hence $G_s$ is not an optimal storage graph, which contradicts to the assumption.

In addition, each vertex $V_i$ must be reachable from $V_0$.
Thus, the optimal storage graph $G_s$ for all problems must be a spanning tree $T$ rooted at dummy vertex $V_0$ in graph $G$.
\end{proof}
\fi

\papertext{\vspace{-8pt}}
\noindent
\papertext{Recall that a spanning tree is a tree where every vertex is connected and reachable, and has no cycles.}
For Problems \ref{prob1} and \ref{prob2}, we have the following observations.
\techreport{
A {\em minimum spanning tree} is defined as a spanning tree of smallest weight, where the weight of a tree is the sum of all its edge weights.
}
A {\em shortest path tree} is defined as a spanning tree where the path from root to each vertex
is a shortest path between those two in the original graph: this would be simply consist of the edges
that were explored in an execution of Dijkstra's shortest path algorithm.

\papertext{\vspace{-8pt}}
\begin{lemma} \label{le:prob1}
The optimal storage graph $G_s$ for Problem \ref{prob1} is a minimum spanning tree of $G$ rooted at $V_0$, considering only the weights $\Delta_{i,j}$.
\end{lemma}
\papertext{\vspace{-15pt}}
\begin{lemma} \label{le:prob2}
The optimal storage graph $G_s$ for Problem \ref{prob2} is a shortest path tree of $G$ rooted at $V_0$, considering only the weights $\Phi_{i,j}$.
\end{lemma}

\subsection{ILP Formulation}\label{ssec:ILP}

We present an ILP formulation of the optimization problems described above.
Here, we take Problem \ref{prob6} as an example;
other problems are similar.
Let $x_{i,j}$ be a binary variable for each edge $(V_i,V_j) \in E$,
indicating whether edge $(V_i,V_j)$ is in the storage graph or not.
Specifically, $x_{0,j}=1$ indicates that version $V_j$ is materialized,
while $x_{i,j}=1$ indicates that the modification from version $i$ to version $j$ is stored where $i\neq 0$.
Let $r_i$ be a continuous variable for each vertex $V_i\in V$, where $r_0=0$; $r_i$ captures
the recreation cost for version $i$ (and must be $\le \theta$).

{
\vskip 4pt
\hrule
\vskip 2pt
{\bf minimize} $\Sigma_{(V_i,V_j) \in E} x_{i,j} \times \Delta_{i,j}$, subject to:
\begin{enumerate}
\item \label{con:1} $\sum_{i} x_{i,j}=1, \forall j$
\item \label{con:2} $r_j-r_i \geq \Phi_{i,j}$ if $x_{i,j}=1$
\item \label{con:3} $r_i\leq \theta, \forall i$
\end{enumerate}
\hrule
\vskip 2pt
}
\begin{lemma}
Problem \ref{prob6} is equivalent to the optimization problem described above.
\end{lemma}
\papertext{\vspace{-6pt}}
\if{0}
\begin{proof}
First, constraint~\ref{con:1} indicates that each vertex $V_i$, $1 \leq i\leq n$, has one and only one in coming edge as described in Lemma~\ref{MST_SP}.
Constraint~\ref{con:2} indicates that no cycle exists in $G_	W$.
This can be proven by contradiction: when there exists a cycle $\{V_{k_1},$ $\ldots,V_{k_l},V_{k_1}\}$, we have
\begin{align*}
r_{k_2}-r_{k_1}  & \geq \Phi_{k_1,k_2} \\
\ldots & \ldots  \\
r_{k_1}-r_{k_l} &  \geq \Phi_{k_l,k_1} \\
\Rightarrow  0 & \geq \sum_{i=1}^{l} \Phi_{k_i,k_{(i+1)\%l}}
\end{align*}
Thus, constraint \ref{con:1} and \ref{con:2} ensures the resulting storage graph $G_s$ is a spanning tree.
Constraint \ref{con:3} corresponds to recreation cost constraint in Problem \ref{prob6}, but note that $r_i$ is not necessarily the recreation cost for version $V_i$.

First, the solution to Problem \ref{prob6} fulfils all constraints listed in integer linear programming above by setting $r_i=\mathcal{R}_i$.
Thus, $\mathcal{C}\geq \xi$.
Then, we prove $\mathcal{C}\leq \xi$ by contradiction.
Suppose there exists a solution to the linear programming above such that $\xi<\mathcal{C}$.
According to constraint $\ref{con:2}$, $\ref{con:3}$ and spanning tree property (we let the path from root $V_0$ to $V_j$ be $\{V_{k_1}=V_0,V_{k_2},...V_{k_l},V_{k_{(l+1)}}=V_{j}\}$):
\begin{align*}
r_j & \geq r_{k_{l}}+\Phi_{k_{l},j} \geq ... \geq r_0+\sum_{m=1}^{l} \Phi_{k_m,k_{(m+1)}} \\
\theta  & \geq r_j \\
\Rightarrow \theta & \geq r_0+\sum_{m=1}^{l} \Phi_{k_m,k_{(m+1)}}
\end{align*}
Thus, the recreation cost for each version $V_j$ is fulfilled. Hence, $\mathcal{C}$ is not the minimum storage cost in Problem \ref{prob6}, which contradicts the assumption.
\end{proof}
\fi

Note however that the general form of an
ILP does not permit an if-then statement (as in (2) above).
Instead, we can transform to the general form with the aid of a large constant $C$.
Thus, constraint 2 can be expressed as follows:
$$\Phi_{i,j}+r_i-r_j\leq (1-x_{i,j})\times C$$
Where $C$ is a ``sufficiently large'' constant such that no additional constraint is added to the model. For instance, $C$ here can be set as $2*\theta$. On one hand, if $x_{i,j}=1 \Rightarrow \Phi_{i,j}+r_i-r_j\leq 0$. On the other hand, if $x_{i,j}=0 \Rightarrow \Phi_{i,j}+r_i-r_j\leq C$. Since $C$ is ``sufficiently large'', no additional constraint is added.


\section{Computational Complexity}\label{sec:complexity}
In this section, we study the complexity
of the problems listed in Table~\ref{table:prob} under different application scenarios.

\stitle{Problem~\ref{prob1} and \ref{prob2} Complexity.}
As discussed in Section~\ref{sec:probOverview},
Problem~\ref{prob1} and~\ref{prob2}
can be solved in polynomial time by directly applying
a minimum spanning tree algorithm
(Kruskal's algorithm or
Prim's algorithm for undirected graphs; Edmonds' algorithm~\cite{tarjan77} for directed graphs) and Dijkstra's shortest path algorithm respectively.
Kruskal's algorithm has time complexity $O(E \log V)$,
while Prim's algorithm also has time complexity $O(E \log V)$
when using binary heap for implementing the priority queue, and $O(E+V \log V)$
when using Fibonacci heap for implementing the priority queue.
The running time of Edmonds' algorithm is $O(EV)$ and can be reduced to $O(E + V \log V)$ with faster implementation.
Similarly, Dijkstra's algorithm for constructing the shortest path tree starting from the root
has a time complexity of $O(E \log V)$ via a binary heap-based priority queue
implementation and a time complexity of  $O(E+V \log V)$
via Fibonacci heap-based priority queue implementation.

Next, we'll show that Problem~\ref{prob:minstor-sumrec} and \ref{prob:minstor-maxrec} are NP-hard
even for the special case where $\Delta=\Phi$ and $\Phi$ is symmetric.
This will lead to hardness proofs for the other variants.

\stitle{Triangle Inequality.}
The primary challenge that we encounter while demonstrating
hardness is that our deltas must obey the triangle inequality:
unlike other settings where deltas need not obey real constraints,
since, in our case, deltas represent actual modifications
that can be stored, it must obey additional realistic constraints.
This causes severe complications in proving hardness,
often transforming the proofs from very simple to fairly challenging.

Consider the scenario when $\Delta=\Phi$ and $\Phi$ is symmetric.
We take $\Delta$ as an example.
The triangle inequality, can be stated as follows:
$$|\Delta_{p,q}-\Delta_{q,w}| \leq \Delta_{p,w} \leq \Delta_{p,q}+\Delta_{q,w}$$
$$|\Delta_{p,p}-\Delta_{p,q}| \leq \Delta_{q,q} \leq \Delta_{p,p}+\Delta_{p,q}$$
where $p,q,w \in V$ and $p\neq q \neq w$.
The first inequality states that the ``delta''
between two versions can not exceed the
total ``deltas'' of any two-hop path with the
same starting and ending vertex;
while the second inequality indicates
that the ``delta'' between two versions must be bigger
than one version's full storage cost minus another version's full storage cost.
Since each tuple and modification is recorded explicitly when $\Phi$ is symmetric, it is natural that these two inequalities hold.

\begin{figure}[h!]
  \centering
  \papertext{\vspace{-10pt}}
  \includegraphics[width=1\columnwidth]{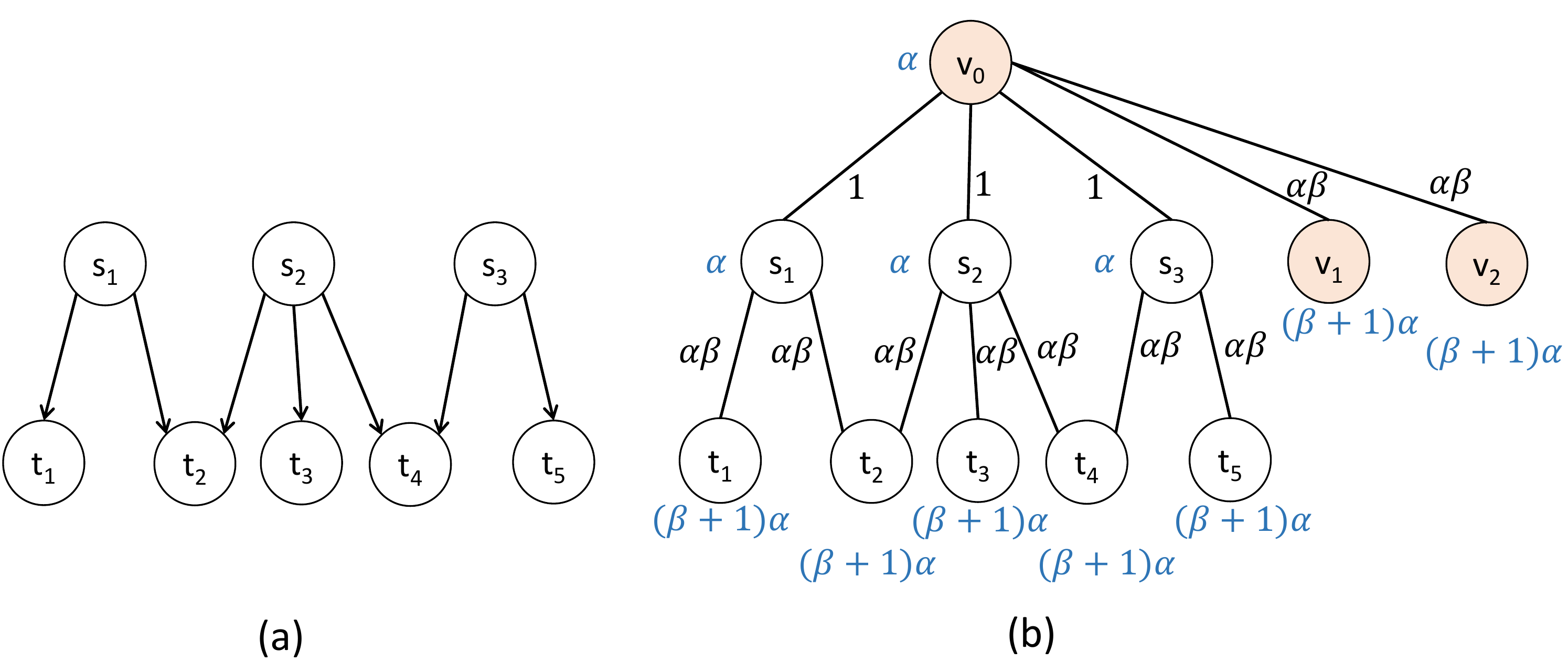}
  \papertext{\vspace{-15pt}}
  \caption{Illustration of Proof of Lemma \ref{prob6:np}}
  \label{fig:l5_proof}
  \papertext{\vspace{-10pt}}
\end{figure}

\stitle{Problem~\ref{prob:minstor-maxrec} Hardness.}
We now demonstrate hardness.
\begin{lemma}\label{prob6:np}
Problem~\ref{prob6}  is NP-hard when $\Delta=\Phi$ and $\Phi$ is symmetric.
\end{lemma}
\papertext{\vspace{-10pt}}
\begin{proof}
Here we prove NP-hardness using a reduction from the set cover problem.
Recall that in the set cover problem,
we are given $m$ sets $S=\{s_1,s_2,...,s_m\}$
and $n$ items $T=\{t_1,t_2,...t_n\}$,
where each set $s_i$ covers some items,
and the goal is to pick $k$ sets $\mathcal{F}\subset S$ such that $\cup_{\{F\in \mathcal{F}\}} F=T$ while minimizing $k$.

Given a set cover instance, we now construct an instance of Problem~\ref{prob6}
that will provide a solution to the original set cover problem.
The threshold we will use in Problem~\ref{prob6} will be $(\beta + 1) \alpha$,
where $\beta, \alpha$ are constants that are each greater than $2 (m + n)$.
(This is just to ensure that they are ``large''.)
We now construct the graph $G(V, E)$ in the following way;
we display the constructed graph in Figure~\ref{fig:l5_proof}.
Our vertex set $V$ is as follows:
\begin{itemize}
\item $\forall s_i\in S$, create a vertex $s_i$ in V.
\item $\forall t_i\in T$, create a vertex $t_i$ in V.
\item create an extra vertex $v_0$, two dummy vertices $v_1,v_2$ in $V$.
\end{itemize}
We add the two dummy vertices simply to ensure that $v_0$ is materialized,
as we will see later. We now define the storage cost for materializing each vertex in $V$
in the following way:
\begin{itemize}
\item $\forall s_i\in S$, the cost is $\alpha$.
\item $\forall t_i\in T$, the cost is $(\beta+1)\alpha$.
\item for vertex $v_0$, the cost is $\alpha$.
\item for vertex $v_1,v_2$, the cost is $(\beta+1)\alpha$.
\end{itemize}
(These are the numbers colored blue in the tree of Figure~\ref{fig:l5_proof}(b).)
As we can see above, we have set the costs in such a way
that the vertex $v_0$ and the vertices corresponding
to sets in $S$ have low materialization cost,
while the other vertices have high materialization
cost: this is by design so that we only end up materializing
these vertices. Our edge set $E$ is now as follows.
\begin{itemize}
\item we connect vertex $v_0$ to each $s_i$ with weight $1$.
\item we connect $v_0$ to both $v_1$ and $v_2$ each with weight $\beta \alpha$.
\item $\forall s_i \in S$, we connect $s_i$ to $t_j$ with weight $\beta \alpha$
when $t_j \in s_i$, where $\alpha=|V|$.
\end{itemize}
It is easy to show that our constructed graph $G$ obeys the triangle inequality.


Consider a solution to Problem~\ref{prob6} on the constructed graph $G$.
We now demonstrate that that solution leads to a solution of the original set
cover problem. Our proof proceeds in four key steps:

\vspace{2pt}
\noindent
{\em Step 1: The vertex $v_0$ will be materialized, while $v_1, v_2$ will not be materialized.}
Assume the contrary---say $v_0$ is not materialized in a solution to Problem~\ref{prob6}.
Then, both $v_1$ and $v_2$ must be materialized,
because if they are not, then the recreation cost of $v_1$ and $v_2$ would
be at least $\alpha (\beta + 1) + 1$, violating the condition of Problem~\ref{prob6}.
However we can avoid materializing $v_1$ and $v_2$,
instead keep the delta to $v_0$ and materialize $v_0$, maintaining the recreation cost as is while reducing the storage cost.
Thus $v_0$ has to be materialized, while $v_1, v_2$ will not be materialized.
(Our reason for introducing $v_1, v_2$ is precisely to ensure
that $v_0$ is materialized so that it can provide basis
for us to store deltas to the sets $s_i$.)

\vspace{2pt}
\noindent
{\em Step 2: None of the $t_i$ will be materialized.}
Say a given $t_i$ is materialized in the solution to Problem~\ref{prob6}.
Then, either we have a set $s_j$ where $s_j$ is connected to $t_i$ in Figure~\ref{fig:l5_proof}(a)
also materialized, or not. Let's consider the former case.
In the former case, we can avoid materializing $t_i$,
and instead add the delta from $s_j$ to $t_i$, thereby reducing
storage cost while keeping recreation cost fixed.
In the latter case, pick any $s_j$ such that $s_j$ is connected to $t_i$
and is not materialized.
Then, we must have the delta from $v_0$ to $s_j$ as part of the solution.
Here, we can replace that edge, and materialized $t_i$, with materialized $s_j$,
and the delta from $s_j$ to $t_i$: this would reduce the total storage cost
while keeping the recreation cost fixed.
Thus, in either case, we can improve the solution if any of the $t_i$ are materialized,
rendering the statement false.

\noindent
{\em Step 3: For each $s_i$, either it is materialized, or the edge from $v_0$ to $s_i$ will
be part of the storage graph.}
This step is easy to see: since none of the $t_i$ are materialized,
either each $s_i$ has to be materialized,
or we must store a delta from $v_0$.

\noindent
{\em Step 4: The sets $s_i$ that are materialized correspond to a minimal set
cover of the original problem.}
It is easy to see that for each $t_j$ we must have an $s_i$
such that $s_i$ covers $t_j$, and $s_i$
is materialized, in order for the recreation cost constraint
to not be violated for $t_j$.
Thus, the materialized $s_i$ must be a set cover for the original problem.
Furthermore, in order for the storage cost to be as small as possible,
as few $s_i$ as possible must be materialized
(this is the only place we can save cost).
Thus, the materialized $s_i$ also correspond to a minimal set cover
for the original problem.

Thus, minimizing the total storage cost is equivalent to minimizing $k$ in set cover problem.
\end{proof}
\techreport{
  Note that while the reduction above uses a graph with only some edge weights (i.e., recreation
  costs of the deltas) known,
  a similar reduction can be derived for a complete graph with all edge weights
  known. Here,
  we simply use the shortest path in the graph reduction above as the edge weight for the missing
  edges. 
  In that case, once again, the storage graph in the solution to Problem~\ref{prob6}
  will be identical to the storage graph described above.
  }

\stitle{Problem~\ref{prob:minstor-sumrec} Hardness:} We now show that
Problem~\ref{prob:minstor-sumrec} is NP-Hard as well.
The general philosophy is similar to the proof in Lemma~\ref{prob6:np}, except that we create $c$ dummy vertices instead of two dummy vertices $v_1,v_2$ in Lemma~\ref{prob6:np}, where $c$ is sufficiently large---this is to once again
ensure that $v_0$ is materialized.
\papertext{The detailed proof can be found in the extended technical report~\cite{tr}.}
\begin{lemma}\label{lemma:prob5} Problem~\ref{prob5} is NP-Hard when $\Delta=\Phi$ and $\Phi$ is symmetric.
\end{lemma}

\techreport{

\begin{figure}[h!]
  \centering
  \papertext{\vspace{-10pt}}
  \includegraphics[width=0.75\columnwidth]{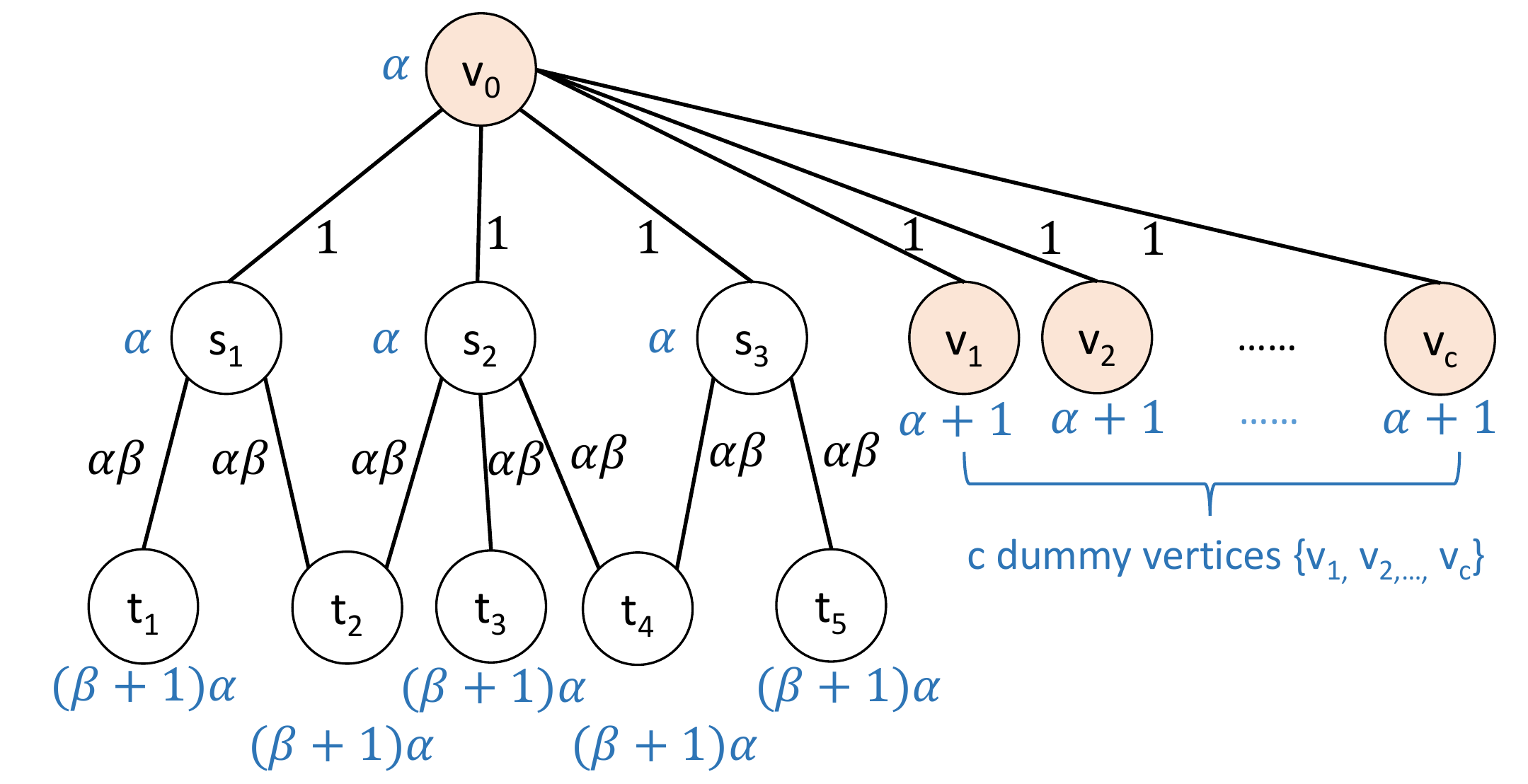}
  \papertext{\vspace{-10pt}}
  \caption{Illustration of Proof of Lemma~\ref{lemma:prob5}  \label{fig:l6_proof}}
  \papertext{\vspace{-10pt}}
\end{figure}

\begin{proof}

We prove NP-hardness using a reduction from the set cover problem.
Recall that in the set cover decision problem,
we are given $m$ sets $S=\{s_1,s_2,...,s_m\}$
and $n$ items $T=\{t_1,t_2,...t_n\}$,
where each set $s_i$ covers some items,
and given a $k$, we ask if
there a subset $\mathcal{F}\subset S$ such that $\cup_{\{F\in \mathcal{F}\}} F=T$ and $|\mathcal{F}|\leq k$.

Given a set cover instance, we now construct an instance of Problem~\ref{prob5}
that will provide a solution to the original set cover decision problem.
The corresponding decision problem for Problem~\ref{prob5} is: 
given threshold  $\alpha+(\beta + 1) \alpha n +k \alpha + (m-k) (\alpha+1)+(\alpha+1) c$ in Problem~\ref{prob5}, is the minimum total storage cost in the constructed graph $G$ no bigger than $\alpha+k \alpha+(m-k)+\alpha \beta n +c$.

We now construct the graph $G(V, E)$ in the following way;
we display the constructed graph in Figure~\ref{fig:l6_proof}.
Our vertex set $V$ is as follows:
\begin{itemize}
\item $\forall s_i\in S$, create a vertex $s_i$ in V.
\item $\forall t_i\in T$, create a vertex $t_i$ in V.
\item create an extra vertex $v_0$, and $c$ dummy vertices $\{v_1,v_2,\dots,v_c\}$ in $V$.
\end{itemize}
We add the $c$ dummy vertices simply to ensure that $v_0$ is materialized,
as we will see later. We now define the storage cost for materializing each vertex in $V$
in the following way:
\begin{itemize}
\item $\forall s_i\in S$, the cost is $\alpha$.
\item $\forall t_i\in T$, the cost is $(\beta+1)\alpha$.
\item for vertex $v_0$, the cost is $\alpha$.
\item for each vertex in $\{v_1,v_2,\dots,v_c\}$, the cost is $\alpha+1$.
\end{itemize}
(These are the numbers colored blue in the tree of Figure~\ref{fig:l6_proof}.)
As we can see above, we have set the costs in such a way
that the vertex $v_0$ and the vertices corresponding
to sets in $S$ have low materialization cost
while the vertices corresponding to $T$ have high materialization
cost: this is by design so that we only end up materializing
these vertices. 
Even though the costs of the dummy vertices is close to that
of $v_0, s_i$, we will show below that they will not be materialized either.
Our edge set $E$ is now as follows.
\begin{itemize}
\item we connect vertex $v_0$ to each $s_i$ with weight $1$.
\item we connect $v_0$ to $v_i, 1 \leq i\leq c$ each with weight $1$.
\item $\forall s_i \in S$, we connect $s_i$ to $t_j$ with weight $\beta \alpha$
when $t_j \in s_i$, where $\alpha=|V|$.
\end{itemize}
It is easy to show that our constructed graph $G$ obeys the triangle inequality.

Consider a solution to Problem~\ref{prob5} on the constructed graph $G$.
We now demonstrate that that solution leads to a solution of the original set
cover problem. Our proof proceeds in four key steps:

\vspace{2pt}
\noindent
{\em Step 1: The vertex $v_0$ will be materialized, while $v_i, 1 \leq i\leq c$ will not be materialized.}
Let's examine the first part of this observation, i.e., that $v_0$ will be materialized.
Assume the contrary.
If $v_0$ is not materialized, then at least one $v_i, 1 \leq i\leq c$, or one of the 
$s_i$ must be materialized,
because if not, then the recreation cost of $\{v_1,v_2, \dots, v_c\}$ would
be at least $(\alpha + 2) c>(\alpha+1) c+\alpha+(\beta + 1) \alpha n +k \alpha + (m-k) (\alpha+1)$, violating the condition (exceeding total recreation cost threshold) of Problem~\ref{prob5}.
However we can avoid materializing this $v_i$ (or $s_i$), instead keep the 
delta from $v_i$ (or $s_i$) to $v_0$ and materialize $v_0$, 
reducing the recreation cost and the storage cost.
Thus $v_0$ has to be materialized. 
Furthermore, since  $v_0$ is materialized, $\forall v_i, 1 \leq i\leq c$ will not be materialized and 
instead we will retain the delta to $v_0$, 
reducing the recreation cost and the storage cost.
Hence, the first step is complete.

\vspace{2pt}
\noindent
{\em Step 2: None of the $t_i$ will be materialized.}
Say a given $t_i$ is materialized in the solution to Problem~\ref{prob5}.
Then, either we have a set $s_j$ where $s_j$ is connected to $t_i$ in Figure~\ref{fig:l6_proof}(a)
also materialized, or not. 
Let us consider the former case.
In the former case, we can avoid materializing $t_i$,
and instead add the delta from $s_j$ to $t_i$, thereby reducing
storage cost while keeping recreation cost fixed.
In the latter case, pick any $s_j$ such that $s_j$ is connected to $t_i$
and is not materialized.
Then, we must have the delta from $v_0$ to $s_j$ as part of the solution.
Here, we can replace that edge, and the materialized $t_i$, with materialized $s_j$,
and the delta from $s_j$ to $t_i$: this would reduce the total storage cost
while keeping the recreation cost fixed.
Thus, in either case, we can improve the solution if any of the $t_i$ are materialized,
rendering the statement false.

\noindent
{\em Step 3: For each $s_i$, either it is materialized, or the edge from $v_0$ to $s_i$ will
be part of the storage graph.}
This step is easy to see: since none of the $t_i$ are materialized,
either each $s_i$ has to be materialized,
or we must store a delta from $v_0$.

\noindent
{\em Step 4: If the minimum total storage cost is no bigger than $\alpha+k \alpha+(m-k)+\alpha \beta n +c$, then there exists a subset $\mathcal{F}\subset S$ such that $\cup_{\{F\in \mathcal{F}\}} F=T$ and $|\mathcal{F}|\leq k$ in the original set cover decision problem, and vice versa.}
Let's examine the first part. If the minimum total storage cost is no bigger than $\alpha+k \alpha+(m-k)+\alpha \beta n +c$, then the storage cost for all $s_i\in S$ must be no bigger than $k \alpha+(m-k)$ since the storage cost for $v_0$, $\{v_1,v_2,\dots,v_c\}$ and $\{t_1,t_2,\dots,t_n\}$ is $\alpha$, $c$ and $\alpha \beta n$ respectively according to Step 1 and 2.
This indicates that at most $k$ $s_i\in S$ is materialized (we let the set of materialized $s_i$ be $M$ and $|M|\leq k$). Next, we prove that each $t_j$ is stored as the modification from the materialized $s_i\in M$. Suppose there exists one or more $t_j$ which is stored as the modification from $s_i\in S-M$, then the total recreation cost must be more than $\alpha+((\beta+1)\alpha n+1)+ k\alpha+(m-k)(\alpha+1)+ (\alpha+1)c$, which exceeds the total recreation threshold. Thus, we have each $t_j\in T$ is stored as the modification from $s_i\in M$.  Let $\mathcal{F}=M$, we can obtain $\cup_{\{F\in \mathcal{F}\}} F=T$ and $|\mathcal{F}| \leq k$. Thus, If the minimum total storage cost is no bigger than $\alpha+k \alpha+(m-k)+\alpha \beta n +c$, then there exists a subset $\mathcal{F}\subset S$ such that $\cup_{\{F\in \mathcal{F}\}} F=T$ and $|\mathcal{F}|\leq k$ in the original set cover decision problem.

Next let's examine the second part. If there exists a subset $\mathcal{F}\subset S$ such that $\cup_{\{F\in \mathcal{F}\}} F=T$ and $|\mathcal{F}|\leq k$ in the original set cover decision problem, then we can materialize each vertex $s_i\in \mathcal{F}$ as well as the extra vertex $v_0$, connect $v_0$ to $\{v_1,v_2,\dots,v_c\}$ as well as $s_j\in S-\mathcal{F}$, and connect $t_j$ to one $s_i\in \mathcal{F}$. The resulting total storage is $\alpha+k \alpha+(m-k)+\alpha \beta n +c$ and the total recreation cost equals to the threshold. Thus, if there exists a subset $\mathcal{F}\subset S$ such that $\cup_{\{F\in \mathcal{F}\}} F=T$ and $|\mathcal{F}|\leq k$ in the original set cover decision problem, then the minimum total storage cost is no bigger than $\alpha+k \alpha+(m-k)+\alpha \beta n +c$.

Thus, the decision problem in Problem ~\ref{prob5} is equivalent to the decision problem in set cover problem.
\end{proof}

\noindent Once again, the problem is still hard if we use a complete graph 
as opposed to a graph where only some edge weights are known.
}

Since Problem~\ref{prob4} swaps the constraint and goal compared to Problem~\ref{prob6},
it is similarly NP-Hard. (Note that the decision versions of the two problems
are in fact identical, and therefore the proof still applies.)
Similarly, Problem~\ref{prob3} is also NP-Hard.
Now that we have proved the NP-hard even in the special case where $\Delta=\Phi$ and $\Phi$ is symmetric, we can conclude that Problem~\ref{prob3},~\ref{prob4},~\ref{prob5},~\ref{prob6}, are NP-hard in a more general setting where $\Phi$ is not symmetric and $\Delta \neq \Phi$, as listed in Table~\ref{table:prob}.

\stitle{Hop-Based Variants.}%
\papertext{
  In the extended technical report,
  we also consider the variant of the problem
  where $\Delta \neq \Phi$ but the recreation cost $\Phi_{ij}$ is the same
  for all pairs of versions, and a version recreation
  cost is simply the number of
  {\em hops} or delta operations to reconstruct the version.
  The reason why this hop-based variant is interesting is because it is related
  to a special case of the {\em d-MinimumSteinerTree} problem,
  namely the {\em d-MinimumSpanningTree} problem,
  i.e., identifying the smallest spanning tree
  where the diameter is bounded by $d$.
  There has been some work on the {\em d-MinimumSpanning\-Tree} problem~\cite{bar2001generalized,charikar1999approximation,kortsarz1997approximating},
  including demonstrating hardness for {\em d-MinimumSpanningTree}
  (using a reduction from SAT),
  and also demonstrating hardness of approximation.%
}%
\techreport{
So far, our focus has been on proving hardness for the special case
where $\Delta = \Phi$ and $\Delta$ is undirected.
We now consider a different kind of special case,
where the recreation cost of all pairs is the same, i.e., $\Phi_{ij} = 1$ for all $i, j$,
while $\Delta \neq \Phi$, and $\Delta$ is undirected.
In this case, we call the recreation cost as the {\em hop cost},
since it is simply the minimum number of delta operations (or "hops")
needed to reconstruct $V_i$.

The reason why we bring up this variant is that this directly corresponds
to a special case of the well-studied {\em d-MinimumSteinerTree} problem:
Given an undirected graph $G=(V,E)$ and a subset $\omega \subseteq V$,
find a tree with minimum weight, spanning the entire vertex subset $\omega$
while the diameter is bounded by $d$.
The special case of {\em d-MinimumSteinerTree} problem when $\omega=V$, i.e., the minimum spanning tree problem with bounded diameter,
directly corresponds
to Problem~\ref{prob:minstor-maxrec} for the hop cost variant we described above.
The hardness for this special case was demonstrated by~\cite{kortsarz1997approximating} using a reduction from the SAT problem:
\begin{lemma} Problem~\ref{prob:minstor-maxrec} is NP-Hard when $\Delta \neq \Phi$ and $\Delta$ is symmetric,
and $\Phi_{ij} = 1$ for all $i, j$.
\end{lemma}
Note that this proof crucially uses the fact
that $\Delta \neq \Phi$
unlike Lemma~\ref{prob6:np} and \ref{lemma:prob5};
thus the proofs are incomparable (i.e.,
one does not subsume the other).

For the hop-based variant, additional results on hardness of approximation
are known by way of the {\em d-MinimumSteinerTree} problem~\cite{bar2001generalized,charikar1999approximation,kortsarz1997approximating}:
\begin{lemma}[\cite{kortsarz1997approximating}] \label{directed}
For any $\epsilon>0$, Problem~\ref{prob6}
has no $\ln n$-$\epsilon$ approximation unless $NP\subset Dtime(n^{\log \log n})$.
\end{lemma}
}%

Since the hop-based variant is a special case of the last column
of Table~\ref{table:prob}, this indicates
that Problem~\ref{prob6} for the most general case is similarly hard
to approximate; we suspect similar results hold for the
other problems as well.
It remains to be seen if hardness of approximation can be demonstrated
for the variants in the second and third last columns.

\section{Proposed Algorithms}
\label{sec:algorithms}
As discussed in Section~\ref{sec:probOverview}, our
different application scenarios lead to different problem formulations,
spanning different constraints
and objectives,
and different assumptions about the nature of $\Phi, \Delta$.

Given that we demonstrated in the previous section
that all the problems are NP-Hard,
we focus on developing efficient heuristics.
In this section, we present two novel heuristics:
first, in Section~\ref{ssec:local_move},
we present {\sc LMG}, or the Local Move Greedy algorithm,
tailored to the case when there is a bound or objective
on the {\em average recreation cost}: thus,
this applies to Problems~\ref{prob:minsumrec} and \ref{prob:minstor-sumrec}.
Second, in Section~\ref{ssec:prim},
we present  {\sc MP}, or Modified Prim's algorithm,
tailored to the case when there is a bound or objective
on the {\em maximum recreation cost}: thus,
this applies to Problems~\ref{prob:minmaxrec} and \ref{prob:minstor-maxrec}.
We present two variants of the {\sc MP} algorithm tailored to
two different settings.


Then, we present two algorithms --- in Section~\ref{subsec:alg_undirected}, we present
an approximation algorithm called {\sc LAST}, and in Section~\ref{subsec:gith}, we present
an algorithm called GitH which is based on Git repack.
Both of these are adapted from literature
to fit our problems and we compare these against our algorithms in Section~\ref{sec:experiments}.
Note that {\sc LAST} does not explicitly optimize any objectives
or constraints in the manner of {\sc LMG}, {\sc MP}, or GitH,
and thus the four algorithms are applicable under different settings;
{\sc LMG} and {\sc MP} are applicable when there is a bound or constraint on the
average or maximum recreation cost, while {\sc LAST} and GitH are applicable
when a ``good enough'' solution is needed.
Furthermore, note that all these algorithms apply to
both directed and undirected versions of the problems,
and to the symmetric and unsymmetric cases.

\papertext{The pseudocodes for the algorithms can be found in
our extended technical report~\cite{tr}.}

\subsection{Local Move Greedy Algorithm}\label{ssec:local_move}

\begin{figure}[t!b]
    \begin{center}
    \includegraphics[width=0.3\textwidth]{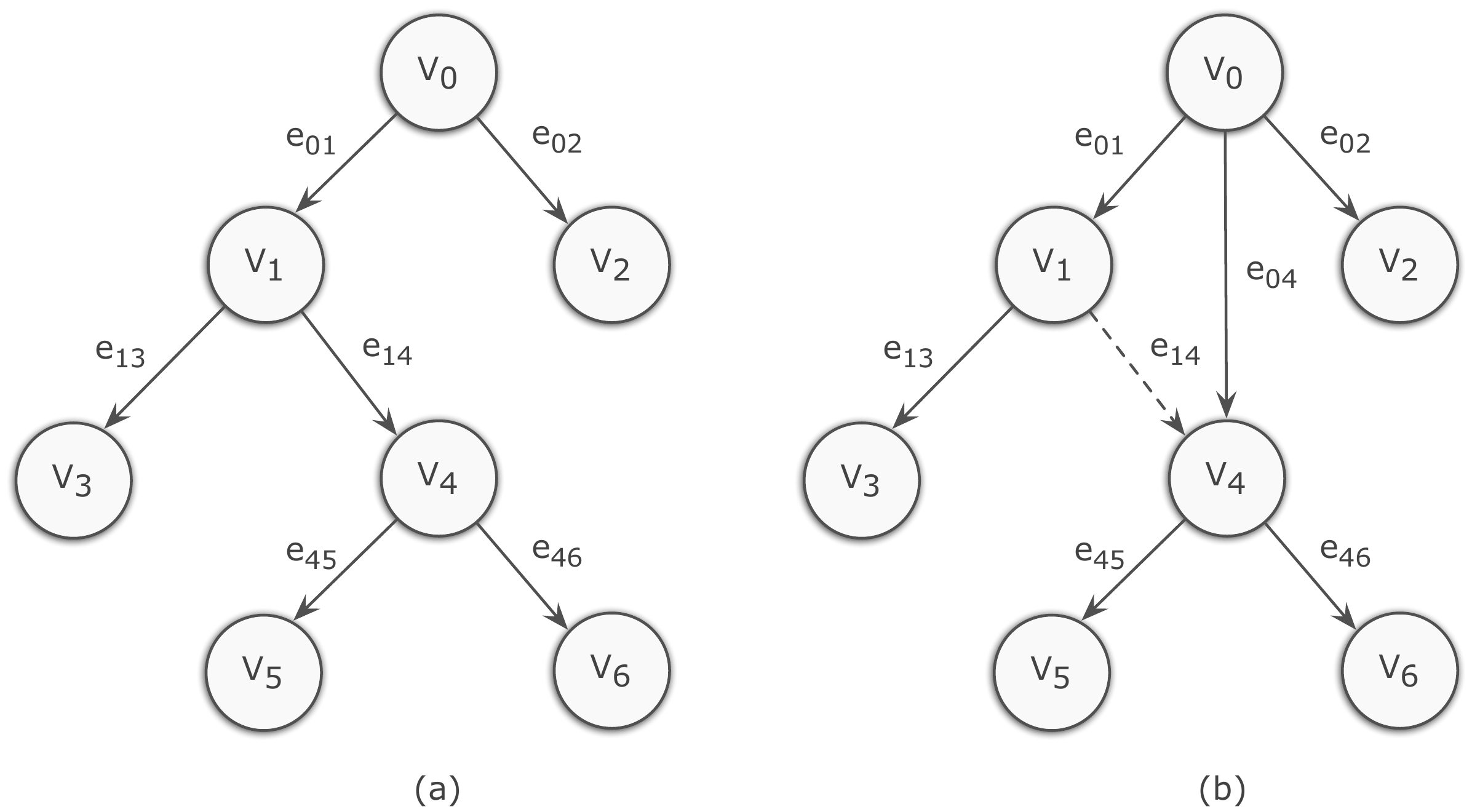}
    \end{center}
    \papertext{\vspace{-15pt}}
    \caption{Illustration of Local Move Greedy Heuristic}
    \papertext{\vspace{-15pt}}
    \label{fig:lmv}
\end{figure}

The LMG algorithm is applicable when we have
a bound or constraint on the average case recreation cost.
We focus on the case where there is a constraint on the storage
cost (Problem~\ref{prob:minsumrec});
the case when there is no such constraint (Problem~\ref{prob:minstor-sumrec})
can be solved by repeated iterations and binary search on the previous problem.

\stitle{Outline.}
At a high level, the algorithm starts with the Minimum Spanning Tree (MST)
as $G_S$, and then greedily adds edges from the Shortest Path Tree (SPT) that are
not present in $G_S$,
while $G_S$ respects the bound on storage cost.


\stitle{Detailed Algorithm.}
The algorithm starts off with $G_S$ equal to the MST.
The SPT naturally contains all the edges corresponding to complete versions.
The basic idea of the algorithm is to replace deltas in $G_S$
with versions from the SPT that maximize the following ratio:
$$ \rho = \frac{\text{reduction in sum of recreation costs}}{\text{increase in storage cost}}$$
This is simply the reduction in total recreation cost per unit addition of weight to the storage graph $G_S$.

Let $\xi$ consists of edges in the SPT not present in the $G_S$
(these precisely correspond to the versions that are not explicitly stored
in the MST, and are instead computed via deltas in the MST).
At each ``round'', we pick the edge $e_{uv} \in \xi$ that maximizes $\rho$,
and replace previous edge $e_{u'v}$ to $v$.
The reduction in the sum of the recreation costs is computed by
adding up the reductions in recreation costs of all $w \in G_S$ that are descendants
of $v$ in the storage graph (including $v$ itself).
On the other hand, the increase in storage cost is simply
the weight of $e_{uv}$ minus the weight of $e_{u'v}$.
This process is repeated as long as the storage budget is not violated.
We explain this with the means of an example.



\begin{example}
Figure~\ref{fig:lmv}(a) denotes the current $G_S$.
Node 0 corresponds to the dummy node.
Now, we are considering replacing edge $e_{14}$
with edge $e_{04}$,
that is, we are replacing a delta to version $5$
with version $5$ itself.
Then, the denominator
of $\rho$ is simply $\Delta_{04}-\Delta_{14}$.
And the numerator is the changes in recreation costs
of versions 4, 5, and 6 (notice that 5 and 6 were
below 4 in the tree.)
This is actually simple to compute:
it is simply three times the change in the recreation cost
of version 4 (since it affects all versions equally).
Thus, we have the numerator of $\rho$
is simply $3 \times (\Phi_{01} + \Phi_{14} - \Phi_{04})$.
\end{example}

\stitle{Complexity.} 
\papertext{Our overall complexity is $O(|V|^2)$. We provide details in
the technical report.}
\techreport{For a given round,
computing $\rho$ for a given edge is $O(|V|)$.
This leads to an overall $O(|V|^3)$ complexity,
since we have up to $|V|$ rounds, and upto $|V|$ edges in $\xi$.
However, if we are smart about this computation
(by precomputing and maintaining across all rounds
the number of nodes ``below'' every node),
we can reduce the complexity of computing $\rho$ for
a given edge to $O(1)$.
This leads to an overall complexity of $O(|V|^2)$
Algorithm~\ref{alg:move-based-balance} provides a pseudocode of the described technique.}

\stitle{Access Frequencies.} Note that the algorithm can easily take
into account access frequencies of different versions
and instead optimize for the total weighted recreation cost (weighted
by access frequencies). The algorithm is similar, except that
the numerator of $\rho$ will capture the
reduction in weighted recreation cost.


\techreport{
\begin{algorithm}[!t]
\DontPrintSemicolon
\SetKwInOut{input}{Input}\SetKwInOut{output}{Output}

{
\input{Minimum Spanning Tree (MST) , Shortest Path Tree (SPT), source vertex $V_0$, space budget $W$}
\output{A tree $T$ with weight $\leq W$ rooted at $V_0$ with minimal sum of access cost}
    Initialize $T$ as MST. \;
    Let $d(V_i)$ be the distance from $V_0$ to $V_i$ in $T$, and $p(V_i)$ denote the parent of $V_i$ in T.
    Let $W(T)$ denote the storage cost of $T$.\;
    \While {$W(T) < W$}{
      $(\rho_{max}, e_{SPT}) \leftarrow (0, \emptyset)$\;
      \ForEach {$e_{uv} \in \xi$}{
	$\mbox{compute } \rho_e$\;
	\If {$\rho_e > \rho_{\max}$}{
	  $(\rho_{max}, \bar{e}) \leftarrow (\rho_e, e_{uv})$\;
	}
      }

	$T \leftarrow T \setminus e_{u^{\prime}v} \cup e_{uv}$; ~~~
	$\xi \leftarrow \xi \setminus e_{uv}$\;
  \If {$\xi = \emptyset$} {
        \Return $T$\;
      }
    }
}
\caption{Local Move Greedy Heuristic}
\label{alg:move-based-balance}
\end{algorithm}
}

\subsection{Modified Prim's Algorithm}\label{ssec:prim} 

Next, we introduce a heuristic algorithm 
based on Prim's algorithm for Minimum Spanning Trees
for Problem~\ref{prob:minstor-maxrec}
where the goal is to reduce total storage cost 
while recreation cost for each version is within threshold $\theta$;
the solution for Problem~\ref{prob:minmaxrec}
is similar.

\stitle{Outline.} At a high level, the algorithm 
is a variant of Prim's algorithm, greedily adding the 
version with smallest storage cost 
and the corresponding edge to form a spanning tree $T$.  
Unlike Prim's algorithm where the spanning tree simply
grows, in this case, even if an edge is present in $T$,
it could be removed in future iterations.
At all stages, the algorithm maintains the invariant that
the recreation cost of all versions in $T$ 
is bounded within $\theta$.

\stitle{Detailed Algorithm.} 
At each iteration, the algorithm picks the version $V_i$
with the smallest storage cost
to be added to the tree.
Once this version $V_i$ is added, we consider adding all deltas
to all other versions $V_j$ such that their recreation cost 
through $V_i$ is within
the constraint $\theta$, and the storage cost does not increase. 
Each version maintains a pair $l(V_i)$ and $d(V_i)$:
$l(V_i)$ denotes the marginal storage cost of $V_i$,
while $d(V_i)$ denotes the total recreation cost of $V_i$.
At the start, $l(V_i)$ is simply the storage cost of $V_i$
in its entirety.

We now describe the algorithm in detail.
Set $X$ represents the current version set 
of the current spanning tree $T$.
Initially $X = \emptyset$.
In each iteration, the version $V_i$ with the
smallest storage cost ($l(V_i)$) 
in the priority queue $PQ$ is picked and added into spanning tree $T$ 
\techreport{(line 7-8)}.
When $V_i$ is added into $T$, we need to update the storage cost and recreation cost for all $V_j$ that are neighbors of $V_i$.
Notice that in Prim's algorithm, we do not need to consider neighbors
that are already in $T$.
However, in our scenario a better path to such a neighbor
may be found and this may result in an update%
\techreport{(line 10-17)}.
For instance, if edge $\langle V_i, V_j \rangle$ 
can make $V_j$'s storage cost 
smaller while the recreation cost for $V_j$ does not increase,
we can update $p(V_j)=V_i$ as well as $d(V_j)$, $l(V_j)$ and $T$.
For neighbors $V_j \not\in T$%
\techreport{(line 19-24)}, 
we update $d(V_j)$, $l(V_j)$,$p(V_j)$ if edge $\langle V_i, V_j \rangle$ can make $V_j$'s storage cost smaller and the recreation cost for $V_j$ is no bigger than $\theta$.
\techreport{Algorithm \ref{alg:prim} terminates in $|V|$ iterations since one version is added into $X$ in each iteration.}

\begin{figure}[t!]
  \begin{minipage}[t]{0.5\linewidth}
    \centering
	 \includegraphics[width=0.8\columnwidth,,height=3cm]{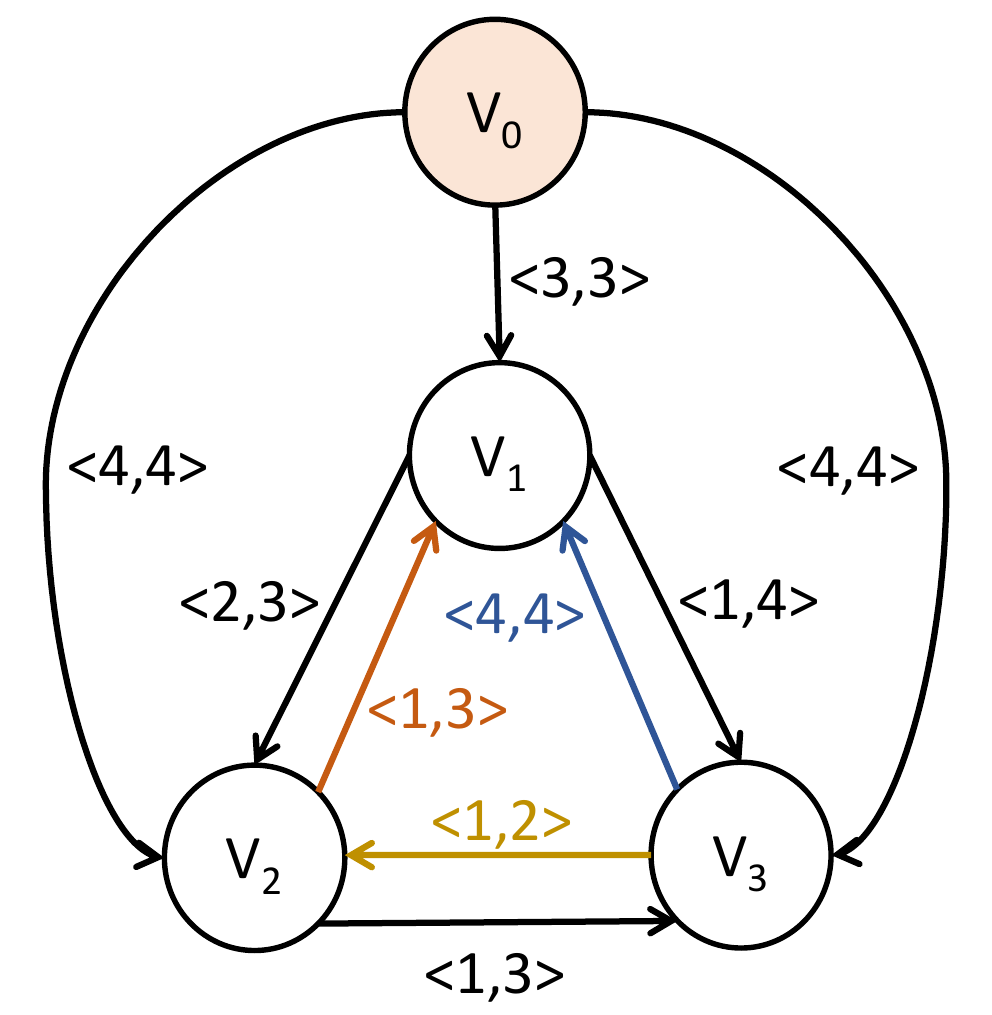}
	 \papertext{\vspace{-10pt}}
	  \caption{Directed Graph $G$}
  	\label{fig:directed}
  \end{minipage}
  \begin{minipage}[t]{0.5\linewidth}
    \centering
	\centering
  \includegraphics[width=0.8\columnwidth,,height=3cm]{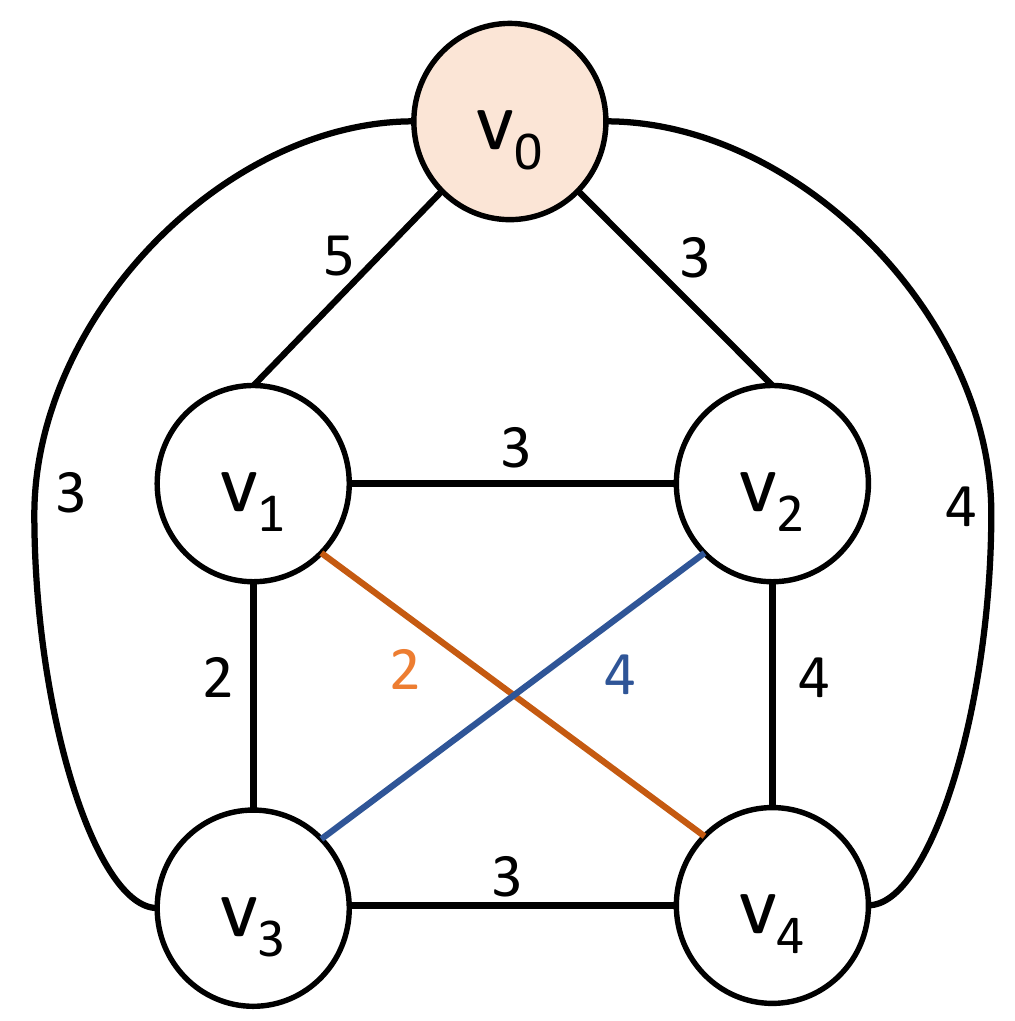}
  \papertext{\vspace{-10pt}}
\caption{ Undirected Graph $G$}
  \label{fig:graph}
  \end{minipage}
    \papertext{\vspace{-5pt}}
\end{figure}

\begin{figure}[t!]
  \centering
  \includegraphics[width=1\columnwidth]{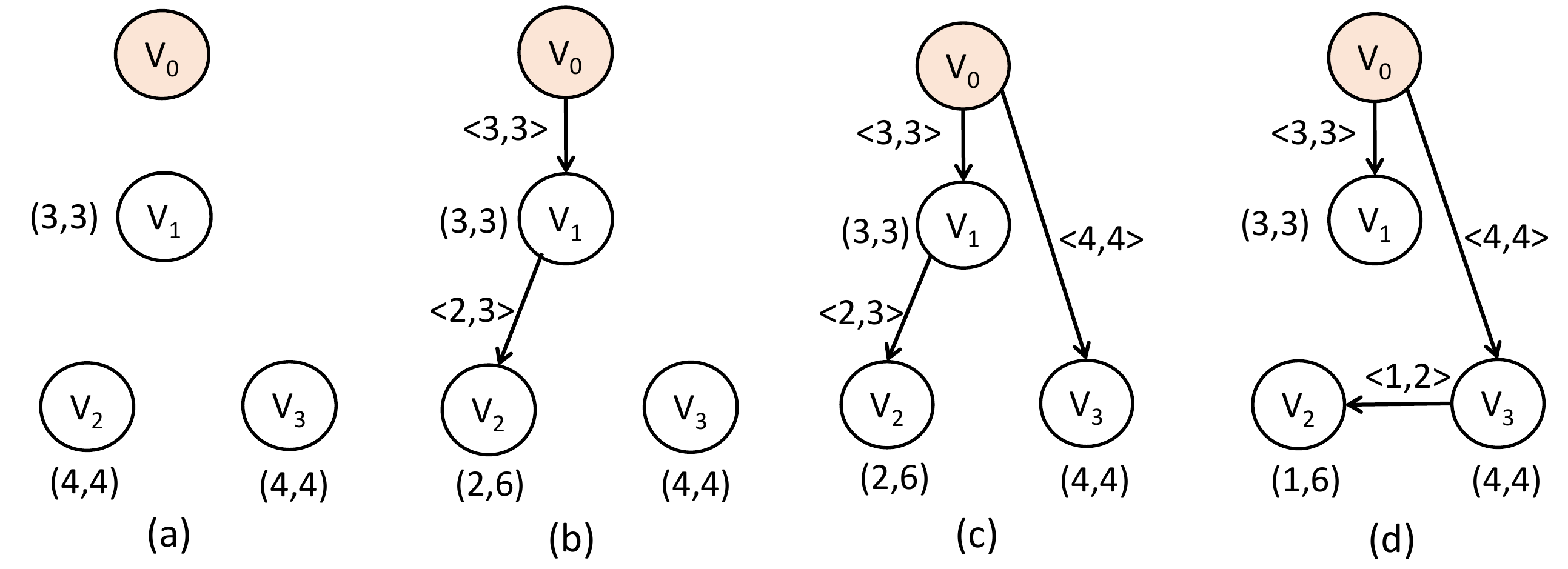}
  \papertext{\vspace{-15pt}}
  \caption{Illustration of Modified Prim's algorithm in Figure \ref{fig:directed}}
  \label{fig:alg}
  \papertext{\vspace{-15pt}}
\end{figure}

\begin{example}
Say we operate on $G$ given by Figure \ref{fig:directed}, and let
the threshold $\theta$ be $6$. 
Each version $V_i$ is associated with a pair $\langle l(V_i),d(V_i)\rangle$. 
Initially version $V_0$ is pushed into priority queue. 
When $V_0$ is dequeued, each neighbor $V_j$ updates $<l(V_j),d(V_j)>$ as shown in Figure \ref{fig:alg} (a). 
Notice that $l(V_i), i \neq 0$ for all $i$ is simply the 
storage cost for that version.
For example, when considering edge $(V_0,V_1)$, 
$l(V_1)=3$ and $d(V_1)=3$ 
is updated since recreation cost 
(if $V_1$ is to be stored in its entirety) 
is smaller than threshold $\theta$, i.e., $3<6$. 
Afterwards, version $V_1,V_2$ and $V_3$ are inserted into the priority queue. 
Next, we dequeue $V_1$ since $l(V_1)$ is smallest among the
versions in the priority queue,
and add $V_1$ to the spanning tree. 
We then update $<l(V_j),d(V_j)>$ for all neighbors of $V_1$, e.g., 
the recreation cost for version $V_2$ will be $6$ 
and the storage cost will be $2$ when considering edge $(V_1, V_2)$. 
Since $6 \leq 6$, $(l(V_2), d(V_2))$ 
is updated to $(2,6)$ as shown in Figure \ref{fig:alg} (b);
however, $<l(V_3),d(V_3)>$ will not be updated 
since the recreation cost is $3+4>6$ when considering edge $(V_1, V_3)$. 
Subsequently, version $V_2$ is dequeued because it has
the lowest $l(V_2)$, and is added to the tree,
giving Figure \ref{fig:alg} (b).
Subsequently, version $V_3$ are dequeued. 
When $V_3$ is dequeued from $PQ$, $(l(V_2), d(V_2))$ is updated. 
This is because the storage cost for $V_2$ can be updated to $1$ 
and the recreation cost is still $6$ when considering edge $(V_3, V_2)$, even if $V_2$ is already in $T$ as shown in Figure  \ref{fig:alg} (c). 
Eventually, we get the final answer in Figure \ref{fig:alg} (d).
\end{example}


\stitle{Complexity.} The complexity of the algorithm 
is the same as that of Prim's algorithm, i.e., $O(|E|\log |V|)$.
\techreport{Each edge is scanned once and the priority queue 
need to be updated once in the worst case.}






\techreport{

\begin{algorithm}[!t]
\SetKwInOut{input}{Input}\SetKwInOut{output}{Output}

{
\input{Graph $G=(V,E)$, threshold $\theta$}
\output{Spanning Tree $T=(V_T,E_T)$}
Let $X$ be the version set of current spanning tree $T$;
 Initially $T=\emptyset, X=\emptyset$\;
 Let $p(V_i)$ be the parent of $V_i$;
 $l(V_i)$ denote the storage cost from $p(V_i)$ to $V_i$,
 $d(V_i)$ denote the recreation cost from root $V_0$ to version $V_i$,

 Initially $\forall i \neq 0$, $d(V_0)=l(V_0)=0, d(V_i)=l(V_i)=\infty$ \;
 Enqueue $<V_0,(l(V_0),d(V_0)) >$ into priority queue $PQ$\;
 ($PQ$ is sorted by $l(v_i)$)\;

\While{$PQ\neq \emptyset $}{
	$<V_i,(l(V_i),d(V_i))> \leftarrow \text{top}(PQ)$, $\text{dequeue}(PQ)$\;
	$T=T\cup<V_i,p(V_i)>$, $X=X\cup V_i$\;
	\For{$V_j\in (V_i$'s neighbors in $G)$ }{ 
	\If{$V_j\in X$}{
		\If{$(\Phi_{i,j}+d(V_i))\leq d(V_j)$ and $\Delta_{i,j}\leq l(V_j)$}{
			$T=T-<V_j,p(V_j)>$\;
			$p(V_j)=V_i$\;
			$T=T\cup<V_j,p(V_j)>$ 
			$d(V_j) \leftarrow \Phi_{i,j}+d(V_i)$\;
			$l(V_j) \leftarrow \Delta_{i,j}$\;
		}
	}
	\Else{
    	    \If{$(\Phi_{i,j}+d(V_i))\leq \theta$ and $\Delta_{i,j}\leq l(V_j)$}{
		$d(V_j) \leftarrow \Phi_{i,j}+d(V_i)$\;
		$l(V_j) \leftarrow \Delta_{i,j}$;
		$p(V_j)=V_i$\;
		enqueue(or update) $<V_j,(l(V_j),d(V_j))>$ in $PQ$\;
	    }
         }

	 }

   }

}

\caption{Modified Prim's Algorithm}
\label{alg:prim}
\end{algorithm}

}

%
%



\subsection{LAST Algorithm}\label{subsec:alg_undirected}

Here, we sketch an algorithm from previous work~\cite{khuller1995balancing}
that enables us to find a tree with a good balance 
of storage and recreation costs, under the
assumptions that $\Delta=\Phi$ and $\Phi$ is symmetric. 

\papertext{
\stitle{Sketch.} The algorithm, which takes a parameter $\alpha$ as input, starts with a minimum spanning tree and does a depth-first traveral (DFS) on it.
When visiting $V_i$ during the traversal, if it finds that the recreation cost for $V_i$
exceeds $\alpha \times$ the cost of the shortest path from $V_0$ to $V_i$, then 
this current path is replaced with the shortest path to the node. 
It can be shown that the total cost of the resulting spanning tree is within  $(1+2/(\alpha-1))$ times the cost of minimum spanning tree in $G$.
Even though the algorithm was proposed for undirected graphs, 
it can be applied to the directed graph case but without any comparable guarantees.
We refer the reader to the full version for more details and pseudocode~\cite{tr}.
}

\techreport{
\stitle{Outline.} The algorithm starts from a minimum spanning tree and does a depth-first traveral (DFS) over the minimum spanning tree. 
During the process of DFS, if the recreation cost for a node exceeds the pre-defined threshold (set up front), then 
this current path is replaced with the shortest path to the node.
}

\techreport{
\stitle{Detailed Algorithm.}
As discussed in Section~\ref{ssec:graph}, 
balancing between recreation cost and storage cost is 
equivalent to balancing between the minimum spanning tree and the shortest path tree rooted at $V_0$.
Khuller et al.~\cite{khuller1995balancing} studied the problem of balancing minimum spanning tree and shortest path tree in an undirected graph,
where the resulting spanning tree $T$ has the following properties,
given parameter $\alpha$:
\begin{itemize}
\item For each node $V_i$: the cost of path from $V_0$ to $V_i$ in $T$ is within $\alpha$ times the shortest path from $V_0$ to $V_i$ in $G$. 
\item The total cost of $T$ is within $(1+2/(\alpha-1))$ times the cost of minimum spanning tree in $G$. 
\end{itemize}
Even though Khuller's algorithm is meant for undirected graphs, 
it can be applied to the directed graph case without any comparable guarantees.
The pseudocode is listed in Algorithm~\ref{alg:MST_SPT}.
}

%
%
%

\techreport{
Let $MST$ denote the minimum spanning tree 
of graph $G$ and $SP(V_0,V_i)$ denote the 
shortest path from $V_0$ to $V_i$ in $G$.
The algorithm starts with the
$MST$ and then conducts a depth-first traversal in $MST$.
Each node $V$ keeps track of its path cost 
from root as well as its parent, 
denoted as $d(V_i)$ and $p(V_i)$ respectively.
Given the approximation parameter $\alpha$, 
when visiting each node $V_i$, we first check whether $d(V_i)$ is bigger than $\alpha\times SP(V_0,V_i)$ where $SP$ stands for shortest path.
If yes, we replace the path to $V_i$ with the shortest path from root to $V_i$ in $G$ and update $d(V_i)$ as well as $p(V_i)$.  In addition, we keep updating $d(V_i)$ and $p(V_i)$ during depth first traversal%
}
\techreport{as stated in line 4-7 of Algorithm~\ref{alg:MST_SPT}.} 

\begin{figure}[t!]
\papertext{\vspace{-10pt}}
  \centering
  \includegraphics[width=0.9\columnwidth,]{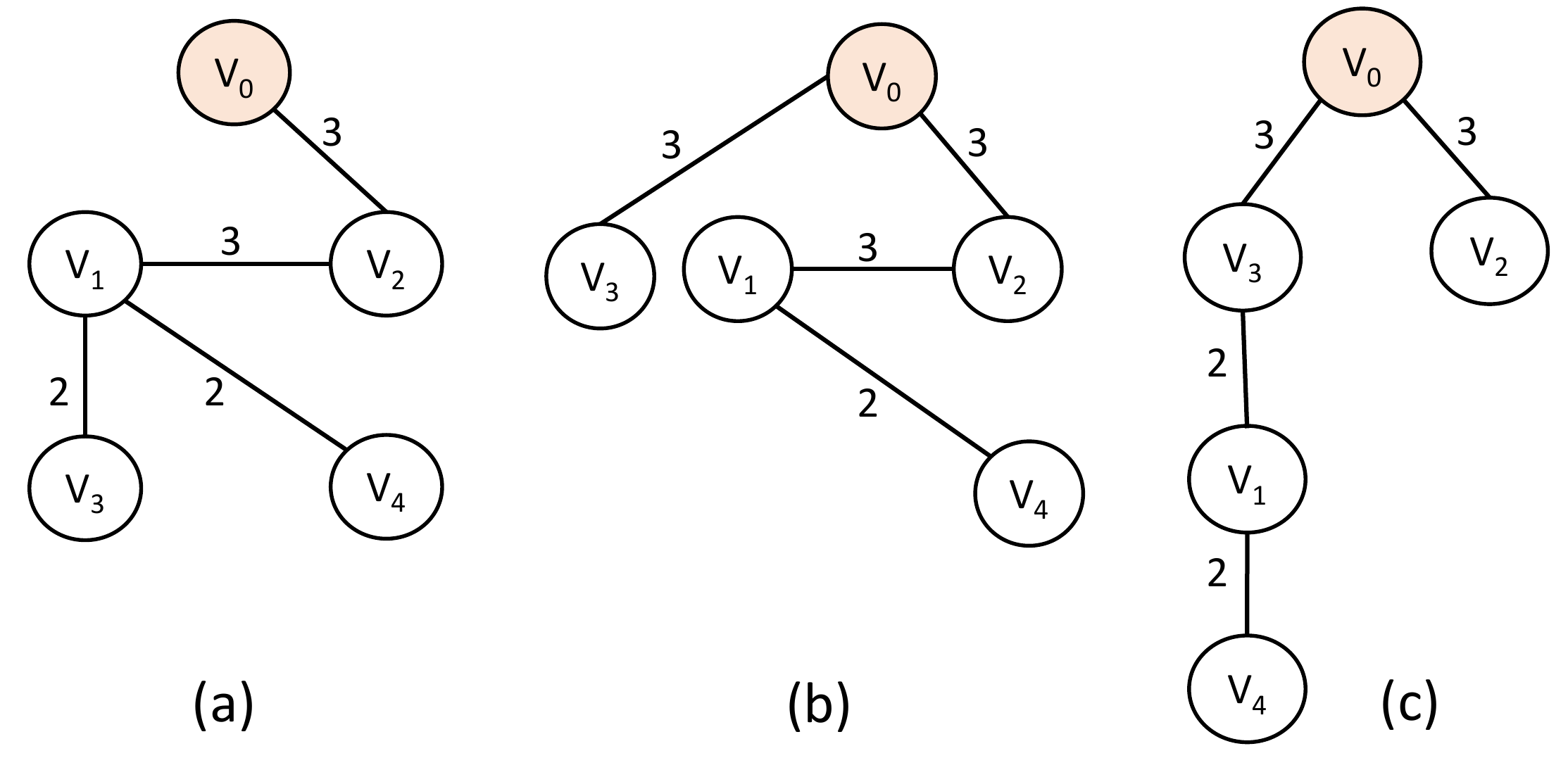}
  \papertext{\vspace{-10pt}}
  \caption{Illustration of LAST on Figure \ref{fig:graph}}
  \label{fig:alg2}
  \papertext{\vspace{-10pt}}
\end{figure}

\begin{example}
Figure~\ref{fig:alg2} (a) is the minimum spanning tree (MST) rooted at node $V_0$ of $G$ in Figure~\ref{fig:graph}. The approximation threshold $\alpha$ is set to be 2. The algorithm starts with the MST and conducts a  depth-first traversal in the MST from root $V_0$.  When visiting node $V_2$, $d(V_2)=3$ and the shortest path to node $V_2$ is $3$, thus $3<2\times 3$. We continue to visit node $V_2$ and $V_3$. When visiting $V_3$, $d(V_3)=8>2\times 3$ where $3$ is the shortest path to $V_3$ in $G$. Thus, $d(V_3)$ is set to be $3$ and $p(V_3)$ is set to be node $0$ by replacing with the shortest path $\langle V_0, V_3\rangle$ as shown in Figure~\ref{fig:alg2} (b). Afterwards, the back-edge $<V_3,V_1>$ is traversed in MST. Since $3+2<6$, where $3$ is the current value of $d(V_3)$, $2$ is the edge weight of $(V_3,V_1)$ and $6$ is the current value in $d(V_1)$, thus $d(V_1)$ is updated as 5 and $p(V_1)$ is updated as node $V_3$. At last node $V_4$ is visited, $d(V_4)$ is first updated as $7$%
\techreport{according to line 3-7}. 
Since $7<2\times 4$, lines 9-11 are not executed. Figure~\ref{fig:alg2} (c) is the resulting spanning tree of the algorithm, 
where the recreation cost for each node is under the constraint and the total storage cost is $3+3+2+2=10$.
\end{example}

\techreport{
\begin{algorithm}[!t]

\SetKwInOut{input}{Input}\SetKwInOut{output}{Output}

{
\input{Graph $G=(V,E)$, $MST$, $SP$}
\output{Spanning Tree $T=(V_T,E_T)$}

Initialize $T$ as $MST$.\
Let $d(V_i)$ be the distance from $V_0$ to $V_i$ in $T$ and $p(V_i)$ be the parent of $V_i$ in $T$.\

\While{DFS traversal on $MST$}{
	$(V_i,V_j) \leftarrow$ the edge currently in traversal\;
    \If{$d(V_j)>d(V_i)+e_{i,j}$}{
	$d(V_j) \leftarrow (d(V_i)+e_{i,j})$\;
	$p(V_j) \leftarrow V_i$\;
	}
	\If{$d(V_j)>\alpha*SP(V_0,V_j)$}{
    add shortest path $(V_0,V_j)$ into $T$\;
    $d(V_j) \leftarrow SP(V_0,V_j)$\;
    $p(V_j) \leftarrow V_0$\;
}

}

}

\caption{Balance MST and Shortest Path Tree~\cite{khuller1995balancing}}
\label{alg:MST_SPT}
\end{algorithm}
}

\stitle{Complexity.}
The complexity of the algorithm is $O(|E|\log |V|)$.
\techreport{Given the minimum spanning tree and shortest path tree rooted at $V_0$, Algorithm~\ref{alg:MST_SPT} is conducted via depth first traversal on MST. It is easy to show that the complexity for Algorithm~\ref{alg:MST_SPT} is $O(|V|)$. The time complexity for computing minimum spanning tree and shortest path tree is $O(|E|\log |V|)$ using heap-based priority queue. 
}
\papertext{Further details can be found in the technical report.}

\subsection{Git Heuristic}\label{subsec:gith}
This heuristic is an adaptation of the current heuristic used by Git and we refer to it as GitH.
\papertext{We sketch the algorithm here and refer the reader to the extended version for more details~\cite{tr}.}
\techreport{We sketch the algorithm here and refer the reader to Appendix~\ref{sec:git-repack} for our analysis of Git's heuristic.}
GitH uses two parameters: $w$ (window size) and $d$ (max depth).


We consider the versions in an non-increasing order of their sizes. The first version in this ordering is chosen as
the root of the storage graph and has depth $0$ (i.e., it is materialized).  At all times, we maintain a sliding
window containing at most $w$ versions. For each version $V_i$ after the first one, let
$V_l$ denote a version in the current window. We compute:
$\Delta'_{l,i} = \Delta_{l,i}/(d - d_l)$, where $d_l$ is the depth of $V_l$ (thus deltas
with shallow depths are preferred over slightly smaller deltas with higher depths). We find the version $V_j$
with the lowest value of this quantity and choose it as $V_i$'s parent (as long as $d_j < d$). The depth of
$V_i$ is then set to $d_j + 1$. The sliding window is modified to move $V_l$ to the end of the window (so it
will stay in the window longer), $V_j$ is added to the window, and the version at the beginning of the window
is dropped.

%
%
%

\stitle{Complexity.}
The running time of the heuristic is $O(|V|\log|V| + w|V|)$, excluding the time to construct deltas.
\begin{figure*}
\begin{tabular}{|l||r|r|r|r|}
\hline
\bfseries Dataset & DC & LC & BF & LF \\ \hhline{|=||=|=|=|=|}
Number of versions &  100010 & 100002 & 986 & 100 \\ \hline
Number of deltas &  18086876 & 2916768 & 442492 & 3562 \\ \hline
Average version size (MB) & 347.65 & 356.46 & 0.401 & 422.79 \\ \hline
MCA-Storage Cost (GB)  & 1265.34 & 982.27 & 0.0250 & 2.2402 \\ \hline
MCA-Sum Recreation Cost (GB) & 11506437.83 & 29934960.95 & 0.9648 & 47.6046 \\ \hline
MCA-Max Recreation Cost (GB) & 257.6 & 717.5 & 0.0063 & 0.5998 \\ \hline
SPT-Storage Cost (GB) &  33953.84 & 34811.14 & 0.3854 & 41.2881 \\ \hline
SPT-Sum Recreation Cost (GB)  & 33953.84 & 34811.14 & 0.3854 & 41.2881 \\ \hline
SPT-Max Recreation Cost (GB) &  0.524 & 0.55 & 0.0063 & 0.5091 \\ \hline
\end{tabular}
\includegraphics[height=0.125\textwidth, width=0.35\textwidth,trim=0 200 0 0]{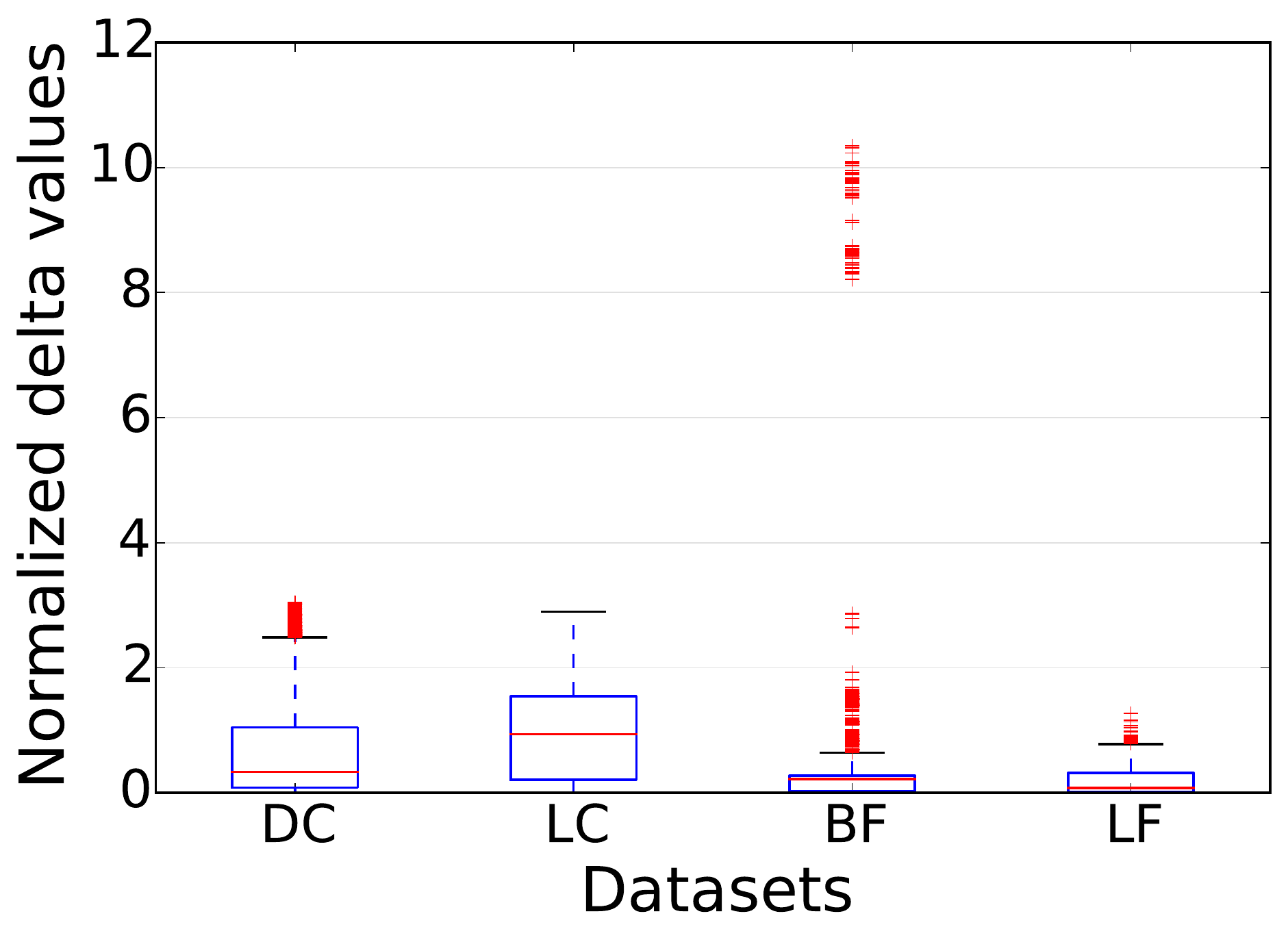}
\caption{Dataset properties and distribution of delta sizes (each delta size scaled by the average version size in the dataset).}
\label{table:datasets}
\papertext{\vspace{-10pt}}
\end{figure*}

\section{Experiments}
\label{sec:experiments}
We have built a prototype version management system, that will serve as a foundation to \dhub~\cite{bhardwaj2014datahub}.
The system provides a subset of Git/SVN-like interface for dataset versioning. 
Users interact with the version management system in a client-server model over HTTP.
The server is implemented in Java, and is responsible for storing the version history of the repository as well as the actual files in them.
The client is implemented in Python and provides functionality to create (commit) and check out versions of datasets, and create and merge branches.
Note that, unlike traditional VCS which make a best effort to perform automatic merges, in our system we let the user perform the
merge and notify the system by creating a version with more than one parent.

\stitle{Implementation.}
In the following sections, we present an extensive evaluation of our designed algorithms using a combination of synthetic and derived real-world datasets.
Apart from implementing the algorithms described above, LMG and LAST require both SPT and MST as input.
For both directed and undirected graphs, we use Dijkstra's algorithm to find the single-source shortest path tree (SPT).
We use Prim's algorithm to find the minimum spanning tree for undirected graphs.
For directed graphs, we use an implementation~\cite{mca-impl} of the Edmonds' algorithm~\cite{tarjan77} for computing the min-cost arborescence (MCA).
We ran all our experiments on a 2.2GHz Intel Xeon CPU E5-2430 server with 64GB of memory, running 64-bit Red Hat Enterprise Linux 6.5.

\subsection{Datasets}
We use four data sets: two synthetic and two derived from real-world source code repositories. 
Although there are many publicly available source code repositories with large numbers of commits (e.g., in {\tt GitHub}),
those repositories typically contain fairly small (source code) files, and further the changes between versions tend
to be localized and are typically very small; we expect dataset versions generated during collaborative data
analysis to contain much larger datasets and to exhibit large changes between versions. We were unable to find any
realistic workloads of that kind.

Hence, we generated realistic dataset versioning workloads as follows. First, we wrote a {\em synthetic version generator
suite}, driven by a small set of parameters, that is able to generate a variety of version histories and corresponding datasets.
Second, we created two real-world datasets using publicly available forks of popular repositories on {\tt GitHub}.
We describe each of the two below.


\vskip 2pt
\noindent
\textbf{Synthetic Datasets:}
Our synthetic dataset generation suite\footnote{Our synthetic dataset generator may be
of independent interest to researchers working on version management.} takes a two-step approach to generate a dataset that we sketch below.
The first step is to generate a version graph with the desired structure, controlled
by the following parameters:
\begin{itemize}
\item
\texttt{number of commits,} i.e., the total number of versions. 
\item
\texttt{branch interval and probability,} the number of consecutive versions after which a branch can be created, and probability of creating a branch. 
\item
\texttt{branch limit,} the maximum number of branches from any point in the version history. We choose a number in $[1, $ \texttt{branch limit}$]$ uniformly at random when we decide to create branches.
\item
\texttt{branch length,} the maximum number of commits in any branch. The actual length is a uniformly chosen integer between 1 and \texttt{branch length}.
\end{itemize}

Once a version graph is generated, the second step is to generate the appropriate versions and compute the deltas.
The files in our synthetic dataset are ordered CSV files (containing tabular data) and we use deltas based on UNIX-style diffs.
The previous step also annotates each edge $(u,v)$ in the version graph with edit commands that can be used to produce $v$ from $u$.
Edit commands are a combination of one of the following six instructions -- add/delete a set of consecutive rows, add/remove a column,
and modify a subset of rows/columns. 

Using this, we generated two synthetic datasets (Figure \ref{table:datasets}):
\begin{itemize}
\item
\textbf{Densely Connected (DC):} This dataset is based on a ``flat'' version history, i.e., number of branches is high, they occur often and have short lengths.
For each version in this data set, we compute the delta with all versions in a 10-hop distance in the version graph to populate additional entries in $\Delta$ and $\Phi$.
\item
\textbf{Linear Chain (LC):} This dataset is based on a ``mostly-linear'' version history, i.e., number of branches is low, they occur after large intervals and have longer lenghts.
For each version in this data set, we compute the delta with all versions within a 25-hop distance in the version graph to populate $\Delta$ and $\Phi$.
\end{itemize}
%

\noindent
\textbf{Real-world datasets:}
We use 986 forks of the Twitter Bootstrap repository and 100 forks of the Linux repository, to derive our real-world workloads.
For each repository, we checkout the latest version in each fork and concatenate all files in it (by traversing the directory structure in lexicographic order).
Thereafter, we compute deltas between all pairs of versions in a repository, provided the size difference between the versions under consideration is less than a threshold.
We set this threshold to 100KB for the Twitter Bootstrap repository and 10MB for the Linux repository.
This gives us two real-world datasets, Bootstrap Forks (BF) and Linux Forks (LF), with properties shown in Figure~\ref{table:datasets}.

%
%

\subsection{Comparison with SVN and Git}
We begin with evaluating the performance of two 
popular version control systems, SVN (v1.8.8) and Git (v1.7.1), using the LF dataset. 
We create an FSFS-type repository in SVN, which is more space efficient that a Berkeley DB-based repository~\cite{svn-fsfs}. We then 
import the entire LF dataset into the repository in a single commit. The amount of space occupied by
the \texttt{db/revs/} directory is around 8.5GB and it takes around 48 minutes to complete the import.
We contrast this with the naive approach of applying a \texttt{gzip} on the files which results in 
total compressed storage of 10.2GB. In case of Git, we add and commit the files in the repository and then run a
\texttt{git repack -a -d --depth=50 --window=50} on the repository\footnote{Unlike \texttt{git repack}, 
\texttt{svnadmin pack} has a negligible effect on the storage cost as it primarily aims to
reduce disk seeks and per-version disk usage penalty by concatenating files into a single
``pack''~\cite{svnpackbook, svnpacknotes}.}. The size of the Git pack file is 202 MB
although the repack consumes 55GB memory and takes 114 minutes (for higher window sizes, Git fails to
complete the repack as it runs out of memory). 

In comparison, the solution found by the MCA algorithm occupies 516MB of
compressed storage (2.24GB when uncompressed) when using UNIX \texttt{diff} for computing the deltas. 
To make a fair comparison with Git, we use \texttt{xdiff} from the LibXDiff
library~\cite{libxdiff} for computing the deltas, which forms the basis of Git's delta computing
routine. Using \texttt{xdiff} brings down the total storage cost to
just 159 MB. The total time taken is around 102 minutes; this includes the time taken to compute the
deltas and then to find the MCA for the corresponding graph. 

\begin{figure*}[t!bp]
    \papertext{\vspace{-10pt}}
  \centering
  \papertext{\vspace{-10pt}}
    \includegraphics[width=0.9\textwidth, keepaspectratio]{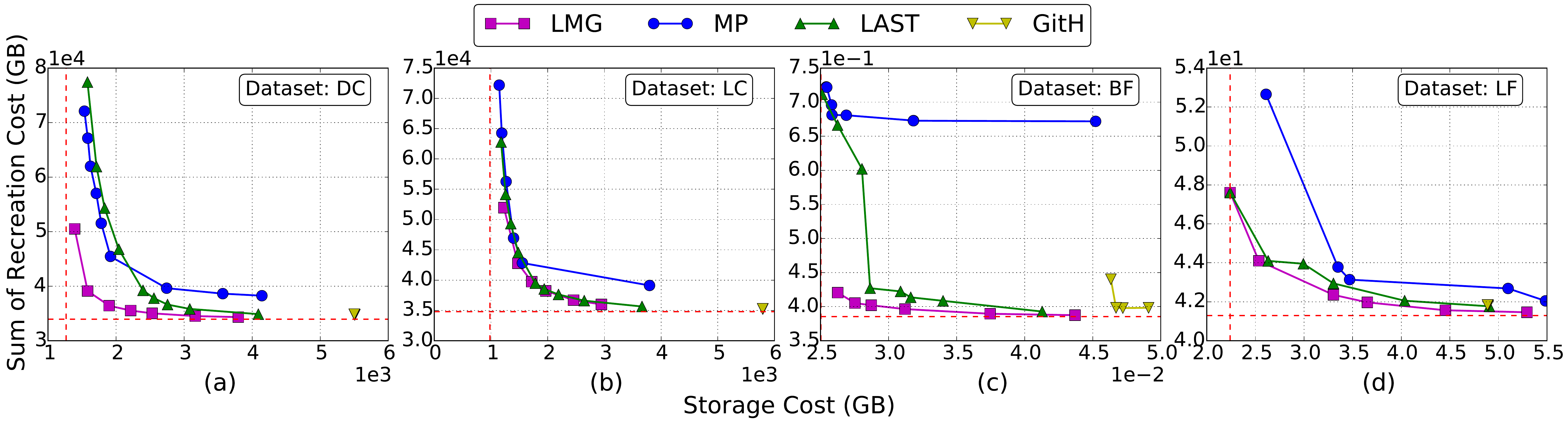}
    \papertext{\vspace{-10pt}}
\caption{Results for the directed case, comparing the storage costs and total recreation costs} 
\label{fig:dir_results_sum}
\papertext{\vspace{-10pt}}
\end{figure*}

The main reason behind SVN's poor performance is its use of ``skip-deltas'' to ensure that at most $O(\log n)$ deltas are needed for reconstructing
any version~\cite{skipdeltas}; that tends to lead it to repeatedly store redundant delta information as a result of
which the total space requirement increases significantly. 
The heuristic used by Git is much better than SVN (Section~\ref{subsec:gith}). 
However as we show later (Fig.~\ref{fig:dir_results_sum}), our implementation of that heuristic
(GitH) required more storage than LMG for guaranteeing similar
recreation costs.

\begin{figure}[t!bp]
  \centering
  \papertext{\vspace{-10pt}}
    \includegraphics[width=0.45\textwidth]{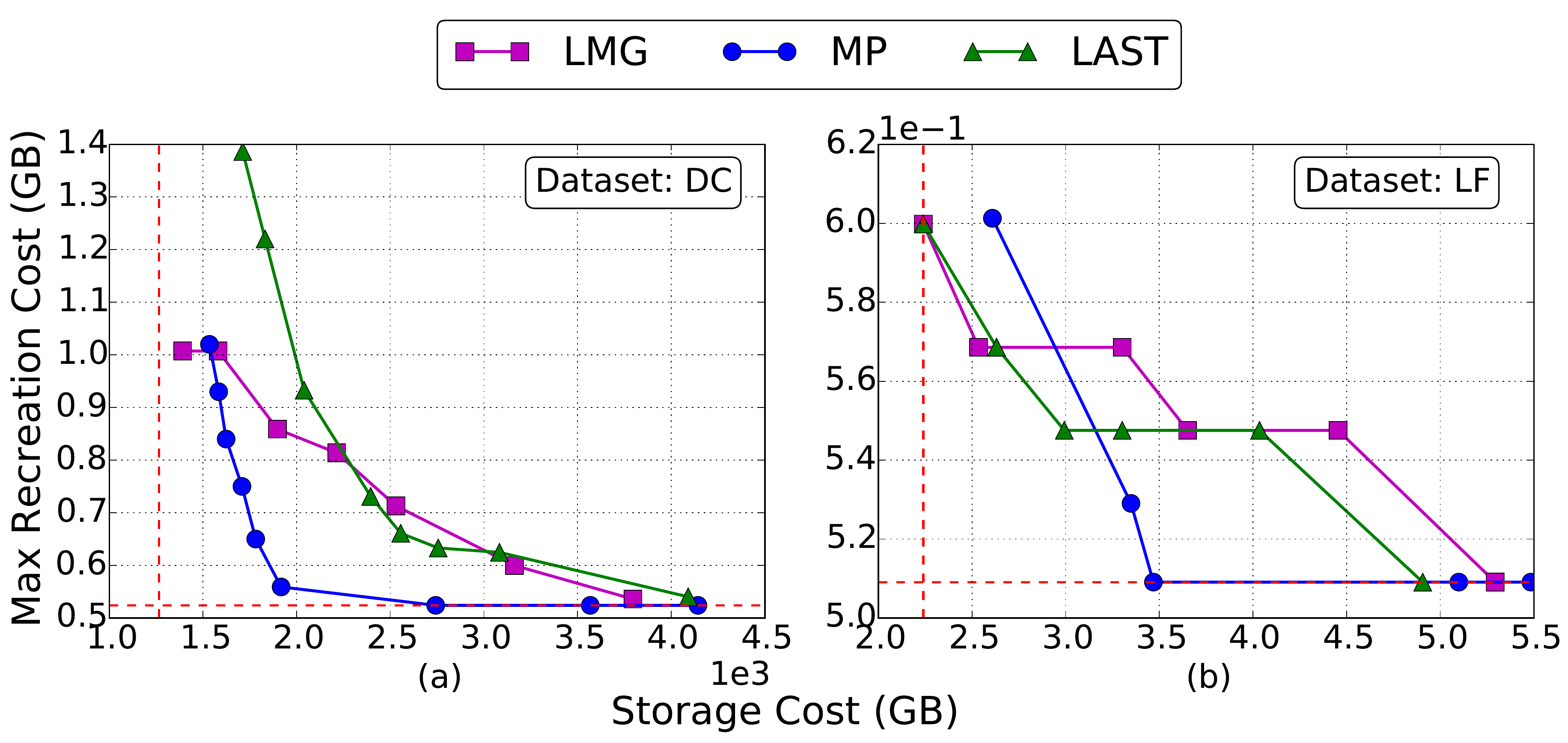}
    \papertext{\vspace{-12pt}}
\caption{Results for the directed case, comparing the storage costs and maximum recreation costs} 
\label{fig:dir_results_max}
\papertext{\vspace{-15pt}}

\end{figure}

\begin{figure*}
  \centering
    \includegraphics[width=0.9\textwidth]{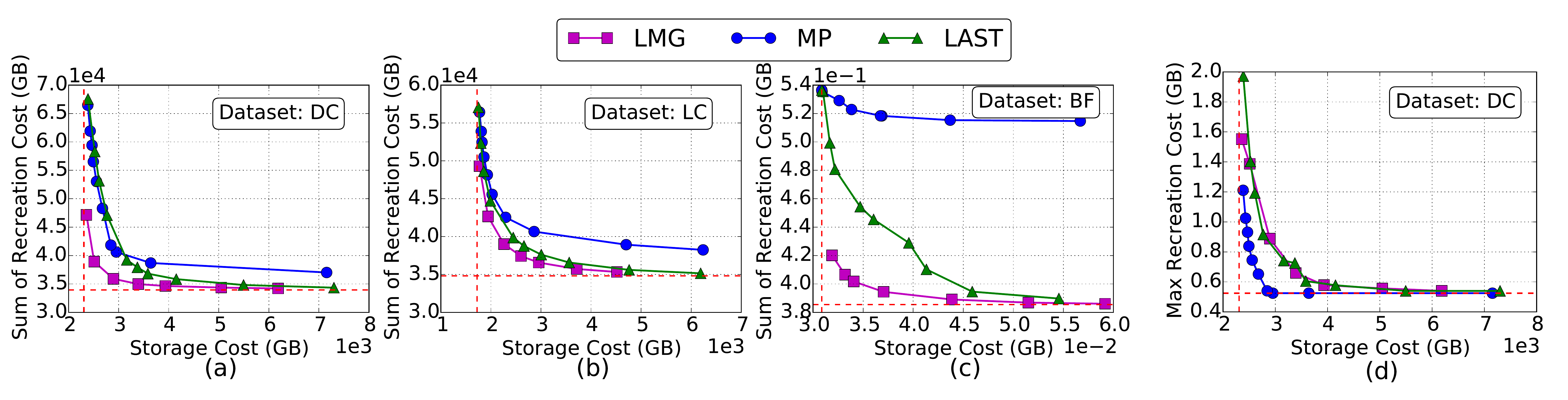}
    \papertext{\vspace{-10pt}}
\caption{Results for the undirected case, comparing the storage costs and total recreation costs (a--c) or maximum recreation costs (d)} 
\papertext{\vspace{-10pt}}
\label{fig:undir_results}
\end{figure*}

\subsection{Experimental Results}

\stitle{Directed Graphs.}
\label{sec:dirgraphs}
We begin with a comprehensive evaluation of the three algorithms, LMG, MP, and LAST, on directed datasets.
Given that all of these algorithms have parameters that can be used to trade off the storage cost
and the total recreation cost, we compare them by plotting the different solutions they are able
to find for the different values of their respective input parameters. Figure~\ref{fig:dir_results_sum}(a--d)
show four such plots; we run each of the algorithms with a range of different values for its input
parameter and plot the storage cost and the total (sum) recreation cost for each of the solutions found.
We also show the minimum possible values for these two costs: the vertical dashed red line indicates the
minimum storage cost required for storing the versions in the dataset as found by MCA, and the horizontal
one indicates the minimum total recreation cost as found by SPT (equal to the sum of all version
sizes).

The first key observation we make is that,
the total recreation cost decreases drastically by allowing a small increase in the storage budget over MCA.
For example, for the DC dataset, the sum recreation cost for MCA is over 11 PB (see Table~\ref{table:datasets})
as compared to just 34TB for the SPT solution (which is the minimum possible). As we can see from Figure~\ref{fig:dir_results_sum}(a),
a space budget of 1.1$\times$ the MCA storage cost reduces the sum of recreation cost by three orders of magnitude.
Similar trends can be observed for the remaining datasets and across all the algorithms.
We observe that LMG results in the best tradeoff between the sum of recreation
cost and storage cost with LAST performing fairly closely.
{\bf An important takeaway here, especially given the amount of prior work that has focused purely on storage cost minimization (Section
        \ref{sec:related}),
is that: it is possible to construct balanced trees
where the sum of recreation costs can be reduced and brought close to that of SPT while using only a fraction of the  space that SPT needs.}

We also ran GitH heuristic on the all the four datasets with varying window and depth settings. For BF, we ran the algorithm with 
four different window sizes (50, 25, 20, 10) for a fixed depth 10 and provided the GitH algorithm with all the deltas that it 
requested. 
For all other datasets, we ran GitH with an infinite window size but restricted it to choose from deltas that were available to the
other algorithms (i.e., only deltas with sizes below a threshold); as we can see, the solutions found by GitH exhibited very good total recreation
cost, but required significantly higher storage than other algorithms. This is not surprising given that GitH is a greedy heuristic that 
makes choices in a somewhat arbitrary order.
    %

In Figures~\ref{fig:dir_results_max}(a--b), we plot the maximum recreation costs instead of the sum of recreation costs across all versions for
two of the datasets (the other two datasets exhibited similar behavior).
The MP algorithm found the best solutions here for all datasets, and we also observed that LMG and LAST both show plateaus for some datasets
where the maximum recreation cost did not change when the storage budget was increased. This
is not surprising given that the basic MP algorithm tries to optimize for the storage cost given a bound on the maximum recreation cost, whereas
both LMG and LAST focus on minimization of the storage cost and one version with high recreation cost is unlikely to affect that significantly.

\stitle{Undirected Graphs.}
\label{sec:undirgraphs}
We test the three algorithms on the undirected versions of three of the datasets (Figure~\ref{fig:undir_results}).
For DC and LC, undirected deltas between pairs of versions were obtained by concatenating the two directional deltas;
for the BF dataset, we use UNIX \texttt{diff} itself to produce undirected deltas.
Here again we observe that LMG consistently outperforms the other algorithms in terms of finding a
good balance between the storage cost and the sum of recreation costs. MP again shows the best
results when trying to balance the maximum recreation cost and the total storage cost.
Similar results were observed for other datasets but are omitted due to space limitations.

\if{0}
In this experiment, we use Algorithm~\ref{alg:move-based-balance} to find the balanced index whose sum of recreation cost is minimized, given a space budget.
The local move-based algorithm constructs an index that aims to balance the MCA and SPT.
We observe from Fig.~\ref{fig:soac} that the gains in sum of recreation cost by allowing a relatively small increase in space budget is significant.
For every dataset, we find the sum of recreation cost corresponding to different space budget constraints.
The space budget provided as an input to the algorithm is more than the space taken by the MCA but less than that taken by the SPT.
As we increase the space budget, the sum of recreation cost of the resulting indices decreases monotonically.
For the \textit{DC} dataset, a space budget of 1.5$\times$ the space taken by MCA reduces the sum of recreation cost of the balanced index by three orders of magnitude.
When compared to SPT, the balanced index has a sum of recreation cost 1.07$\times$ that of SPT while occupying an order of magnitude less space.
The results are similar for the other synthetically generated dataset \textit{LC}.
For the real world dataset \textit{BF}, the sum of recreation cost is reduced by an order of magnitude when the space budget is set to 1.5$\times$ that of MCA.
The sum of recreation cost is slightly more than that of the SPT (1.01$\times$) although the space occupied by the balanced index is an order of magnitude less than that of SPT.
These results suggest that balancing the two indices namely, the MCA and SPT yields significant benefits.

In this section, we describe the results of the experiments with Algorithm~\ref{} which computes an index that tries to maintain a distance of at most $\alpha$ times the shorest path distance from the root to every vertex while at the same time restricts the weight of the index to at most $\beta$ times the weight of the MST.
As mentioned before, this algorithm is a modification of the LAST algorithm proposed for undirected graphs therefore the guarantees provided by LAST may not always hold here.
We provide the $\alpha$ values as input to the algorithm and observe the recreation cost as well as the weight of the tree for the final index.
From Fig.~\ref{}, we observe that as the $\alpha$ values increase, the total weight of index decreases.
This is because as the distance of the vertices from the root increases, the algorithm is not forced to pick heavy edges from the SPT to satisfy the constraint.
Rather it can use existing edges in MCA because of the relaxed requirment on the distance of the vertices from the root.

We evaluate the performance of the Modified Prim algorithm by comparing it with the storage used by the Local move based heuristic for same values of max recreation cost.
We measure the storage cost for three runs of the Modified Prim algorithm with $\theta$ values taken from the result of three runs of the local move based heuristic, namely,  LMB1.5, LMB2 and LMB3.
\fi

\begin{figure}[t]
\papertext{\vspace{-5pt}}
  \centering
    \includegraphics[width=0.45\textwidth]{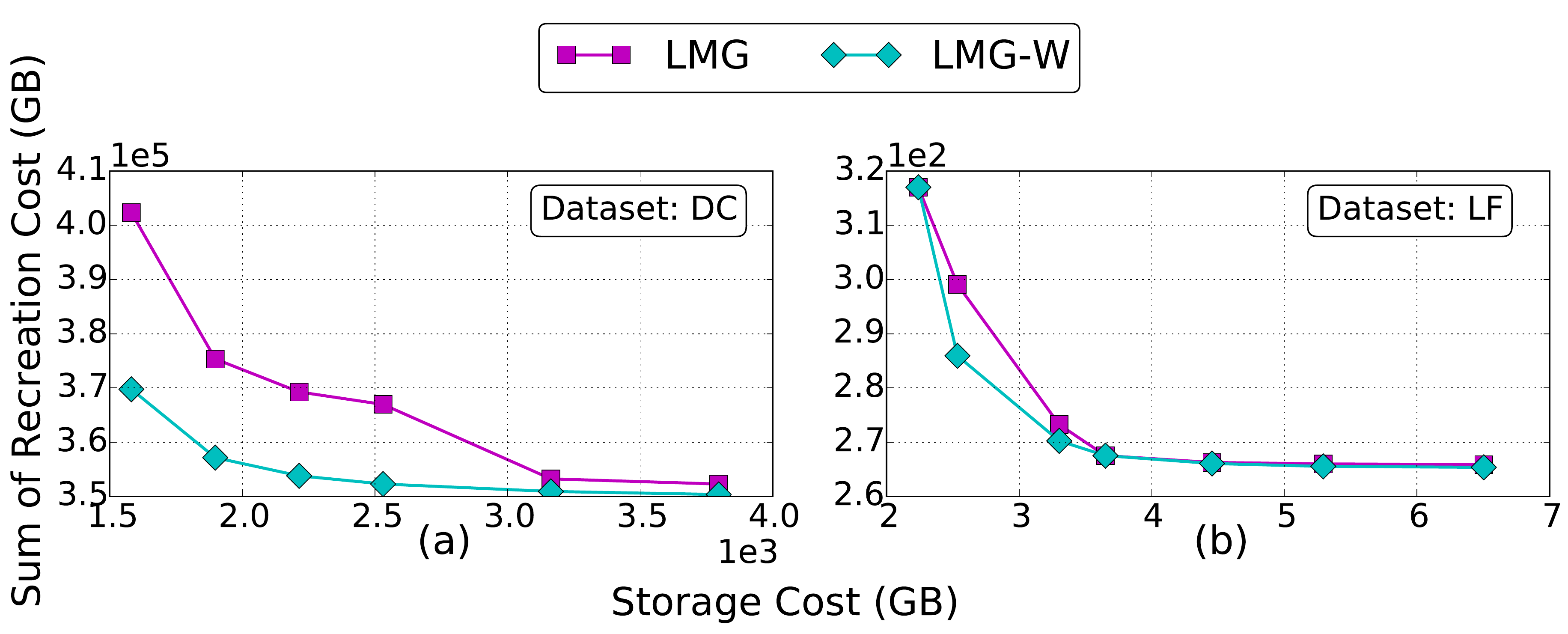}
    \papertext{\vspace{-10pt}}
\caption{Taking workload into account leads to better solutions}
\label{fig:dir_workload}
\papertext{\vspace{-18pt}}
\end{figure}

\stitle{Workload-aware Sum of Recreation Cost Optimization.}
In many cases, we may be able to estimate access frequencies for the various
versions (from historical access patterns), and if available, we may want
to take those into account when constructing the storage graph. The LMG algorithm
can be easily adapted to take such information into account, whereas it is not
clear how to adapt either LAST or MP in a similar fashion.
In this experiment, we use LMG to compute a storage graph such that the sum of
recreation costs is minimal given a space budget, while taking workload
information into account. The worload here assigns a frequency of access
to each version in the repository using a Zipfian distribution
(with exponent 2); real-world access frequencies are known to follow such
distributions.
Given the workload information, the algorithm should find a
storage graph that has the sum of recreation cost less than the index when
the workload information is not taken into account (i.e., all versions are
assumed to be accessed equally frequently). Figure~\ref{fig:dir_workload} shows
the results for this experiment. As we can see, for the DC dataset, taking
into account the access frequencies during optimization led to much better
solutions than ignoring the access frequencies. On the other hand, for the LF
dataset, we did not observe a large difference.

\begin{figure}[t]
\papertext{\vspace{-5pt}}
  \centering
    \includegraphics[width=0.5\textwidth]{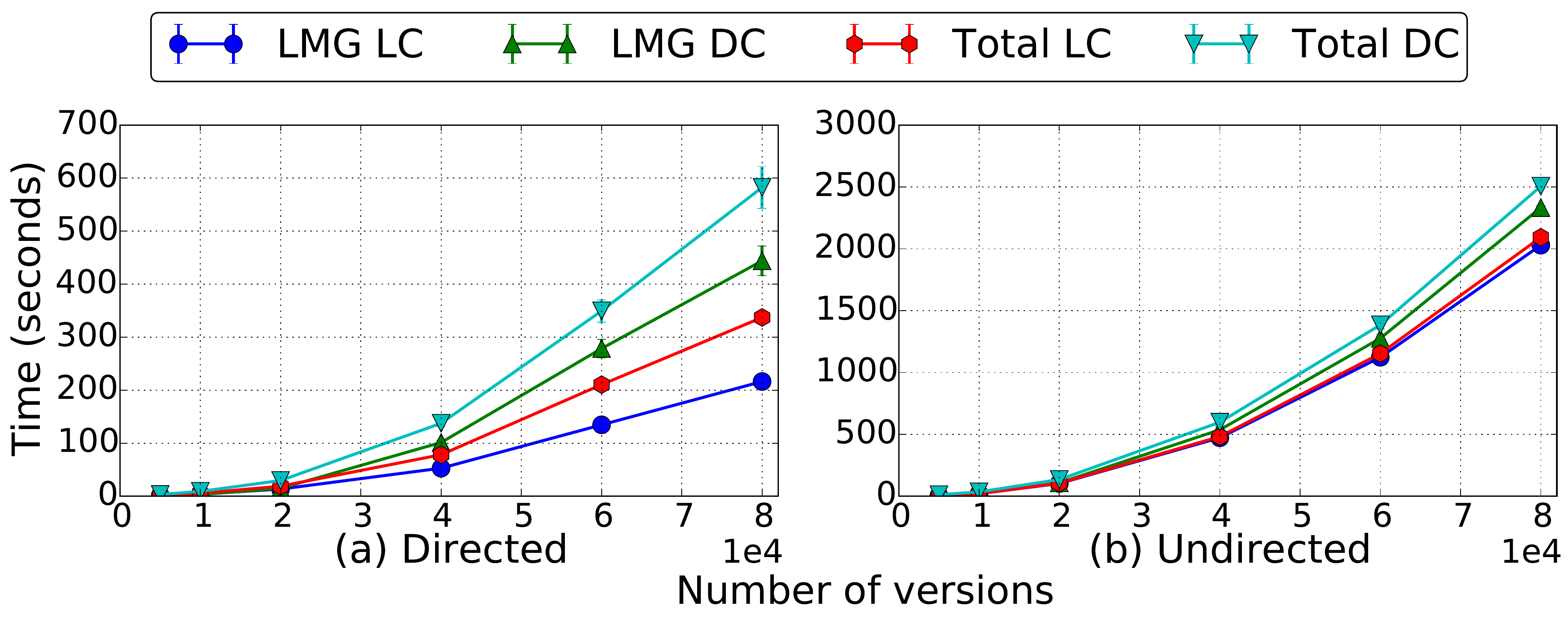}
    \papertext{\vspace{-20pt}}
\caption{Running times of LMG}
\label{fig:runtime_lmg}
\papertext{\vspace{-20pt}}
\end{figure}

\stitle{Running Times.} Here we evaluate the running times of the LMG algorithm. 
Recall that LMG takes MST (or MCA) and SPT as inputs. 
%
In Fig.~\ref{fig:runtime_lmg}, we report the total running time as well as the time taken by LMG
itself. We generated a set of version graphs as subsets of the graphs for LC and DC datasets as follows:
for a given number of versions $n$, we randomly choose a node and traverse the graph starting at that
node in breadth-first manner till we construct a subgraph with $n$ versions.
We generate 5 such subgraphs for increasing values of $n$ and report the average running time for LMG; the
storage budget for LMG is set to three times of the space required by the MST (all our reported
experiments with LMG use less storage budget than that). The time
taken by LMG on DC dataset is more than LC for the same number of versions; 
this is because DC has lower delta values than LC (see
Fig.~\ref{table:datasets}) and thus requires more edges from SPT to satisfy the storage budget. 

On the other hand, MP takes between 1 to 8 seconds on those datasets, when the recreation cost 
is set to maximum. Similar to LMG, LAST requires the MST/MCA and SPT as inputs; however the running 
time of LAST itself is linear and it takes less than 1 second in all cases.
Finally the time taken by GitH on LC and DC datasets, on varying
window sizes range from 35 seconds (window = 1000) to a little more than 120 minutes (window =
100000); note that, this excludes the time for constructing the deltas.

In summary, although LMG is inherently a more expensive algorithm than MP or LAST, it runs in 
reasonable time on large input sizes; we note that all of these times are likely to be dwarfed 
by the time it takes to construct deltas even for moderately-sized datasets.

\stitle{Comparison with ILP solutions.}
\label{exp:ilp}
Finally, we compare the quality of the solutions found by MP with the optimal solution found
using the Gurobi Optimizer for Problem \ref{prob6}.
We use the ILP formulation from Section~\ref{ssec:ILP} with constraint on the maximum recreation cost ($\theta$), and compare the optimal storage cost with that of the MP algorithm (which resulted in solutions with lowest maximum recreation costs in our evaluation).
We use our synthetic dataset generation suite to generate three small datasets, with 15, 25 and 50 versions denoted by v15, v25 and v50 respectively and compute deltas between all pairs of versions.
Table~\ref{table:ilp} reports the results of this experiment, across five $\theta$ values.
The ILP turned out to be very difficult to solve, even for the very small problem sizes, and
in many cases, the optimizer did not finish and the reported numbers are the best solutions found
by it.

As we can see, the solutions found by MP are quite close to the ILP solutions for the small
problem sizes for which we could get any solutions out of the optimizer. However, extrapolating
from the (admittedly limited) data points, we expect that on large problem sizes, MP may be
significantly worse than optimal for some variations on the problems (we note that the optimization
problem formulations involving max recreation cost are likely to turn out to be harder than the
formulations that focus on the average recreation cost). Development of better heuristics and
approximation algorithms with provable guarantees for the various problems that we introduce are
rich areas for further research.

\begin{table}[t]
\centering
\begin{tabular}{|c|c||r|r|r|r|r|}
\hline
& & \multicolumn{5}{|c|}{Storage Cost (GB)} \\ \hhline{|=|=#=|=|=|=|=|}
v15 & $\theta$ & 0.20 & 0.21 & 0.22 & 0.23 & 0.24 \\ \hline
 & ILP & 0.36 & 0.36 & 0.22 & 0.22 & 0.22 \\ \hline
 & MP & 0.36 & 0.36 & 0.23 & 0.23 & 0.23 \\ \hhline{|=|=#=|=|=|=|=|}
v25 & $\theta$ & 0.63 & 0.66 & 0.69 & 0.72 & 0.75 \\ \hline
 & ILP & 2.39 & 1.95 & 1.50 & 1.18 & 1.06 \\ \hline
 & MP & 2.88 & 2.13 & 1.7 & 1.18 & 1.18 \\ \hhline{|=|=#=|=|=|=|=|}
v50 & $\theta$ & 0.30 & 0.34 & 0.41 & 0.54 & 0.68 \\ \hline
 & ILP & 1.43 & 1.10 & 0.83 & 0.66 & 0.60 \\ \hline
 & MP & 1.59 & 1.45 & 1.06 & 0.91 & 0.82 \\ \hline
\end{tabular}
\caption{Comparing ILP and MP solutions for small datasets, given a bound on max recreation cost, $\theta$ (in GB)}
\papertext{\vspace{-16pt}}
\label{table:ilp}
\end{table}

\section{Related Work}
\label{sec:related}

Perhaps the most closely related prior work is source code version systems like Git, Mercurial, SVN, and others, that are widely used
for managing source code repositories. Despite their popularity, these systems largely use fairly simple algorithms underneath that
are optimized to work with modest-sized source code files and their on-disk structures are optimized to work with line-based diffs.
These systems are known to have significant limitations when handling large files and large numbers of versions~\cite{git-comment-fb}.
As a result, a variety of extensions like git-annex~\cite{git-annex}, git-bigfiles~\cite{git-bigfiles}, etc., have been developed
to make them work reasonably well with large files.

There is much prior work in the temporal databases
literature~\cite{Bolour92,DBLP:conf/sigmod/SnodgrassA85,Ozsoyoglu1995,Tansel1993} on managing a linear chain of versions, and retrieving a version as of a specific time
point (called {\em snapshot} queries)~\cite{Salzberg1999}.
\cite{buneman2004archiving} proposed an archiving technique where all versions of the data are merged into one hierarchy.
\techreport{
An element appearing in multiple versions is stored only once along with a timestamp.
This technique of storing versions is in contrast with techniques where retrieval of certain versions may require undoing the changes (unrolling the deltas).
The hierarchical data and the resulting archive is represented in XML format which enables use of XML tools such as an XML compressor for compressing the archive.
It was not, however, a full-fledged version control system representing an arbitrarily graph of versions; rather it focused on algorithms for compactly encoding a linear chain of versions.
}
\papertext{
It was not, however, a full-fledged version control system representing an arbitrarily graph of versions. 
}

Snapshot queries have recently also been studied in
the context of array databases~\cite{soroushB13,seering-versions} and graph databases~\cite{conf/icde/KhuranaD13}.
\papertext{
Seering et al.~\cite{seering-versions} proposed an MST-like technique for storing an arbitrary tree of versions
in the context of scientific databases. 
}
\techreport{
Seering et al.~\cite{seering-versions}  considered the problem of storing an arbitrary tree of
versions in the context of scientific databases; their proposed techniques are based on finding a minimum
spanning tree (as we discussed earlier, that solution represents one extreme in the spectrum of solutions that needs to be considered). 
}
They also proposed several heuristics for choosing which versions to materialize given
the distribution of access frequencies to historical versions.
Several databases support ``time travel'' features (e.g., Oracle Flashback, Postgres~\cite{stonebraker1991postgres}).
However, those do not allow for  branching trees of versions. 
\cite{GatterbauerS11} articulates a similar vision to our overall \dhub vision; however, 
they do not propose formalisms or algorithms to solve the underlying data management challenges.
\techreport{In addition, the schema of tables encoded with Flashback cannot change.}

There is also much prior work on compactly encoding differences between two
files or strings in order to reduce communication or storage costs.
In addition to standard utilities like UNIX {\tt diff}, many sophisticated
techniques have been proposed for computing differences or edit sequences
between two files (e.g., xdelta~\cite{xdelta}, vdelta~\cite{vdelta}, vcdiff~\cite{vcdiff},
zdelta~\cite{trendafilov02zdelta}). That work is largely orthogonal and complementary to our work.

Many prior efforts have looked at the problem of minimizing the total storage
cost for storing a collection of related files (i.e., Problem 1). These works
do not typically consider the recreation cost or the tradeoffs between the two.
Quinlan et al.~\cite{venti} propose an archival ``deduplication'' storage system that identifies
duplicate blocks across files and only stores them once for reducing storage requirements.
Zhu et al.~\cite{zhu} present several optimizations on the basic theme.
Douglis et al.~\cite{Douglis2003} present several techniques to identify pairs of files
that could be efficiently stored using delta compression even if there is no explicit
derivation information known about the two files; similar techniques could be used to
better identify which entries of the matrices $\Delta$ and $\Phi$ to reveal in our
scenario. 
\techreport{
Ouyang et al.~\cite{ouyang2002cluster} studied the problem of compressing a
large collection of related files by performing a sequence of
pairwise delta compressions. They proposed a suite of text clustering
techniques to prune the graph of all pairwise delta encodings and find the
optimal branching (i.e., MCA) that minimizes the total weight. 
}
Burns and Long~\cite{burns1998} present a technique
for in-place re-construction of delta-compressed files using a graph-theoretic approach. That
work could be incorporated into our overall framework to reduce the memory requirements
during reconstruction. 
\techreport{
Similar
dictionary-based reference encoding techniques have been used by~\cite{chan1999cache} to
efficiently represent a target web page in terms of additions/modifications to
a small number of reference web pages. 
Kulkarni et al.~\cite{kulkarni2004} present a more general technique that
combines several different techniques to identify similar blocks among a collection files,
and use delta compression to reduce the total storage cost (ignoring the recreation costs).
}
We refer the reader to a recent survey~\cite{paulostoragesurvey} for a more comprehensive coverage of
this line of work.

\section{Conclusions and Future Work}
Large datasets and collaborative and iterative analysis are becoming a norm
in many application domains; however we lack the data management infrastructure
to efficiently manage such datasets, their versions over time, and derived
data products. Given the high overlap and duplication among the datasets, it 
is attractive to consider using delta compression to store the datasets in a
compact manner, where some datasets or versions are stored as modifications from
other datasets; such delta compression however leads to higher latencies while 
retrieving specific datasets. In this paper, we studied the trade-off between the 
storage and recreation costs in a principled manner, by formulating several 
optimization problems that trade off these two in different ways and showing that
most variations are NP-Hard. We also presented several efficient algorithms that
are effective at exploring this trade-off, and we presented an extensive 
experimental evaluation using a prototype version management system that we have
built. There are many interesting and rich avenues for future work that we 
are planning to pursue. In particular, we plan to develop online algorithms
for making the optimization decisions as new datasets or versions are 
being created, and also adaptive algorithms that reevaluate the optimization
decisions based on changing workload information. We also plan to explore
the challenges in extending our work to a distributed and decentralized
setting.

{
    \papertext{\scriptsize}
\bibliographystyle{abbrv}

}

\appendix
\section{Git repack}
\label{sec:git-repack}

Git uses delta compression to reduce the amount of storage required to store a large number of files (objects) that contain duplicated information.
However, git's algorithm for doing so is not clearly described anywhere.
An old discussion with Linus has a sketch of the algorithm~\cite{linus-repack}.
However there have been several changes to the heuristics used that don't appear to be documented anywhere.

The following describes our understanding of the algorithm based on the latest git source code
\footnote{Cloned from \url{https://github.com/git/git} on 5/11/2015, commit id: 8440f74997cf7958c7e8ec853f590828085049b8}.

Here we focus on ``repack'', where the decisions are made for a large group of objects.
However, the same algorithm appears to be used for normal commits as well.
Most of the algorithm code is in file: \texttt{builtin/pack-objects.c}

\stitle{Step 1:}
Sort the objects, first by ``type'', then by ``name hash'', and then by ``size'' (in the decreasing order).
The comparator is (line 1503):
\begin{verbatim}
static int type_size_sort(const void *_a, const 
     void *_b)
\end{verbatim}
Note the name hash is not a true hash; the \texttt{pack\_name\_hash()} function (\texttt{pack-objects.h}) simply creates a number from the last 16 non-white space characters, with the last characters counting the most (so all files with the same suffix, e.g., \texttt{.c}, will sort together).

\stitle{Step 2:}
The next key function is \texttt{ll\_find\_deltas()}, which goes over the files in the sorted order.
It maintains a list of $W$ objects ($W$ = window size, default 10) at all times.
For the next object, say $O$, it finds the delta between $O$ and each of the objects, say $B$, in the window; it chooses the the object with the minimum value of:
\texttt{delta(B, O) / (max\_depth - depth of B)}
where \texttt{max\_depth} is a parameter (default 50), and depth of B refers to the length of delta chain between a root and B.

The original algorithm appears to have only used \texttt{delta(B, O)} to make the decision,
but the ``depth bias'' (denominator) was added at a later point to prefer slightly larger deltas with smaller delta chains.
The key lines for the above part:
\begin{itemize}
\item line 1812 (check each object in the window):
\begin{verbatim}
ret = try_delta(n, m, max_depth, &mem_usage);
\end{verbatim}

\item lines 1617-1618 (depth bias):
\begin{verbatim}
max_size = (uint64_t)max_size * (max_depth - 
    src->depth) / (max_depth - ref_depth + 1);
\end{verbatim}

\item line 1678 (compute delta and compare size):
\begin{verbatim}
delta_buf = create_delta(src->index, trg->data, 
    trg_size, &delta_size, max_size);
\end{verbatim}
\end{itemize}

\texttt{create\_delta()} returns non-null only if the new delta being tried is smaller than the current delta (modulo depth bias),
specifically, only if the size of the new delta is less than \texttt{max\_size} argument. Note: lines 1682-1688 appear redundant
given the depth bias calculations.

\stitle{Step 3.}
Originally the window was just the last $W$ objects before the object $O$ under consideration.
However, the current algorithm shuffles the objects in the window based on the choices made.
Specifically, let $b_1, \ldots, b_W$ be the current objects in the window.
Let the object chosen to delta against for $O$ be $b_i$.
Then $b_i$ would be moved to the end of the list, so the new list would be: $[b_1, b_2, \ldots, b_{i-1}, b_{i+1}, \ldots, b_W, O, b_i]$.
Then when we move to the new object after $O$ (say $O'$), we slide the window and so the new window then would be: $[b_2, \ldots, b_{i-1}, b_{i+1}, \ldots, b_W, O, b_i, O']$.
Small detail: the list is actually maintained as a circular buffer so the list doesn't have to be physically ``shifted'' (moving $b_i$ to the end does involve a shift though).
Relevant code here is lines 1854-1861.

Finally we note that git never considers/computes/stores a delta between two objects of different types, and it does the above in a multi-threaded fashion,
by partitioning the work among a given number of threads. Each of the threads operates independently of the others.


\end{document}